%% file: TobarEtAl.tex
\documentclass[vecphys]{svmult}

\usepackage{makeidx}         
\usepackage{graphicx}        
\usepackage{multicol}        
\usepackage[bottom]{footmisc}
\def\OO#1{{\cal O}(c^{-#1})}
\def\be{\begin{equation}}
\def\ee{\end{equation}}
\def\bea{\begin{eqnarray}}
\def\eea{\end{eqnarray}}
\newcommand{\ktilde}{\widetilde{\kappa}}

\makeindex             

\begin{document}

\title*{Rotating Resonator-Oscillator Experiments to Test Lorentz Invariance in Electrodynamics}
\titlerunning{Resonator Tests of Lorentz Invariance} 
\author{Michael E. Tobar\inst{1}, Paul L. Stanwix\inst{1}, Mohamad Susli\inst{1}, Peter Wolf\inst{2,3}, Clayton R. Locke\inst{1} \and Eugene N. Ivanov\inst{1}}
\authorrunning{Tobar, Stanwix, \it{et. al.}} 
\institute{University of Western Australia, School of Physics M013, 35 Stirling Hwy., Crawley 6009 WA, Australia.
\texttt{mike@physics.uwa.edu.au}
\and SYRTE, Observatoire de Paris, 61 Av. de l'Observatoire, 75014 Paris, France. 
\and Bureau International des Poids et Mesures, Pavillon de Breteuil, 92312 S\`evres Cedex, France.} 
%
%
\maketitle

\section{Introduction}
\label{Intro}
The Einstein Equivalence Principle (EEP) is a founding principle of relativity \cite{Will}. 
One of the constituent
elements of EEP is Local Lorentz Invariance (LLI), which postulates
that the outcome of a local experiment is independent of the
velocity and orientation of the apparatus. The central importance of
this postulate has motivated tremendous work to
experimentally test LLI. Also, a number of unification theories
suggest a violation of LLI at some level. However, to
test for violations it is necessary to have an alternative theory to
allow interpretation of experiments \cite{Will}, and many have
been developed \cite{Robertson,MaS,LightLee,Ni,Kosto1,KM}. The
kinematical frameworks (RMS) \cite{Robertson, MaS} postulate a
simple parameterization of the Lorentz transformations with
experiments setting limits on the deviation of those parameters from
their values in special relativity (SR). Because of their simplicity
they have been widely used to interpret
many experiments \cite{Brillet,Wolf,Muller,WolfGRG}. More recently,
a general Lorentz violating extension of the standard model of
particle physics (SME) has been developed \cite{Kosto1} whose
Lagrangian includes all parameterized Lorentz violating terms that
can be formed from known fields.

This work analyses rotating laboratory Lorentz invariance
experiments that compare precisely the resonant frequencies of two high-Q factor (or high finesse) cavity resonators. High stability electromagnetic oscillatory fields are generated by implementing state of the art frequency stabilization systems with the narrow line width of the resonators. Previous non-rotating experiments \cite{Lipa,Muller,Wolf04} relied on the rotation of the Earth
to modulate putative Lorentz violating effects. This is not optimal for two
reasons. Firstly, the sensitivity to Lorentz violations is
proportional to the noise of the oscillators at the
modulation frequency, typically best for periods between 10 and
100 seconds. Secondly, the sensitivity is
proportional to the square root of the number of periods of the modulation signal, therefore
taking a relatively long time to acquire sufficient data. Thus, by
rotating the experiment the data integration rate is
increased and the relevant signals are translated to the 
optimal operating regime \cite{Mike}.

In this work we outline the two most commonly used test theories (RMS and SME) for testing LLI of the photon. Then we develop the general frame-work of applying these test theories to resonator experiments with an emphasis on rotating experiments in the laboratory. We compare the inherent sensitivity factors of common experiments and propose some new configurations. Finally we apply the test theories to the rotating cryogenic experiment at the University of Western Australia, which recently set new limits in both the RMS and SME frameworks \cite{PRL}. Note added: Two other concurrent experiments have set some similar limits \cite{schil, achim}.

\section{Common Test Theories to Characterize Lorentz Invariance}
\label{Test}

The most famous test of LLI (or the constancy of the speed of light) was that conducted by Michelson and Morley in 1887 \cite{MM} with a rotating table and a Michelson interferometer. In actual fact, the theoretical framework used by Michelson and Morley was not a test of LLI, since the concept did not exist at the time, but that of an aether drift. The relative motion of the apparatus through the aether was thought to induce a phase difference between the arms of the interferometer (and hence an interference pattern) depending on the orientation. Thus, as the Earth moved from one end of its orbit to the opposite end, the change in its velocity should be a detectable value. Michelson and Morley found no fringe shifts due to Earth motion around the sun and reported a null result. Since the Michelson Morley experiment, there have been many other types of experiments devised to test the validity of SR and the constancy of light. However, to interpret these experiments one must formulate an alternative test theory, and in this section we outline two of the most commonly used.

\subsection{Robertson, Mansouri, Sexl Framework}
\label{RMSFW}

A simple kinematic test theory that has been widely used is that of Robertson, Mansouri and Sexl (RMS)\cite{Robertson,MaS}, where time standards (``clocks'') and length standards (``rods'') are considered without taking into account their underlying structure. This framework postulates a preferred frame $\Sigma(T, \mathbf{X})$ which satisfies LLI, and a moving frame $S(t,\mathbf{x})$, which does not. The prime candidate for the preferred frame is taken as the Cosmic Microwave Background (CMB), since any anticipated non-symmetries are expected to arise from Planck-scale effects during the creation of the universe. In this framework we analyse the Poynting vector direction of the electromagnetic signal with respect to the velocity of the lab through the CMB.

The normal Lorentz Transformations for a boost in the $x$ direction are expressed in a special form below (where $c$ is the speed of light in the $\Sigma$ frame):
\begin{equation}
dT = \frac{1}{a} \left( dt + \frac{v dx}{c^2}\right); 
dX = \frac{dx}{b} + \frac{v}{a} \left(dt + \frac{v dx}{c^2}\right);
dY = \frac{dy}{d};
dZ = \frac{dz}{d};
\label{RMSeqn}
\end{equation}
Here we take a Taylor expansion for $a$, $b$ and $d$ of the form: $a \approx 1 + \alpha v^2/c^2 + \OO4$;  $b \approx 1 + \beta v^2/c^2 + \OO4$; $d \approx 1 + \delta v^2/c^2 + \OO4$. Recalling that $\gamma = \frac{1}{\sqrt{1-v^2/c^2}}$ from Special Relativity (SR), we see that SR predicts $\alpha = -1/2$ and $\beta = 1/2$. Since SR predicts no contraction in directions orthogonal to a boost, it also predicts that $\delta = 0$.  Thus, the RMS parameterizes a possible Lorentz violation by a deviation of the parameters ($\alpha, \beta, \delta$) from the SR values ($-\frac{1}{2}, \frac{1}{2}, 0$). 

By manipulating equation (\ref{RMSeqn}) to form the infinitesimals in the $S$ frame, we can separate the equation into a boost term $(\beta - \alpha - 1)$, anisotropy term $(\delta -  \beta + \frac{1}{2})$ and time dilation parameter $\alpha + \frac{1}{2}$. Thus, a complete verification of LLI in
the RMS framework \cite{Robertson,MaS} requires a test of (i) the
isotropy of the speed of light (measuring $P_{MM}=\delta -  \beta +
\frac{1}{2}$), a Michelson-Morley (MM) experiment \cite{MM}, (ii)
the boost dependence of the speed of light (measuring $P_{KT}=\beta - \alpha -
1$), a Kennedy-Thorndike (KT) experiment \cite{KT} and (iii) the
time dilation parameter (measuring $P_{IS}=\alpha + \frac{1}{2}$), an
Ives-Stillwell (IS) experiment \cite{IS,Saat}. Rotating experiments may be considered Michelson-Morley experiments and only measure $P_{MM}$, so in this section we restrict ourselves to these types of measurements.

Assuming only a MM type Lorentz violation, and setting $ds^2 = c^2dT^2-dX^2-dY^2-dZ^2 = 0$ in $\Sigma$, and transforming according to Eqn. (\ref{RMSeqn}) we find the coordinate travel time of a light signal in $S$ becomes;
\be
dt={dl\over c}\left( P_{MM}\times {\rm sin}^2\theta{v^2\over c^2}\right)+\OO4
\label{cmbvel}
\ee
where $dl = \sqrt{dx^2+dy^2+dz^2}$ and $\theta$ is the angle between the Poynting vector and the velocity {\bf v} of $S$ in $\Sigma$. For a modern MM experiment that measures the difference frequency between two resonant cavities, the fractional frequency difference may be calculated from (\ref{cmbvel}) in a similar way to \cite{WolfGRG} to give:
\begin{equation}
\frac{\Delta\nu_0}{\nu_0} = \frac{P_{MM}}{2\pi
c^2}\left[\oint{\left(\bf{v}.\hat{\bf{I}}_{a} (\it{q_a})\right)^2
dq_a}-\oint{\left(\bf{v}.\hat{\bf{I}}_{b} (\it{q_b})\right)^2
dq_b}\right] \label{RMSunit}
\end{equation}
Where $\hat{\bf{I}}_{j}(q_j)$ is the unit vector in the direction of light propagation (Poynting vector) of each resonator (labeled by subscripts $a$ and $b$), and $q_j$ is the
variable of integration around the closed path coordinates of the Poynting vector of each resonator. 

\begin{figure}
\begin{center}
\includegraphics[width=6cm]{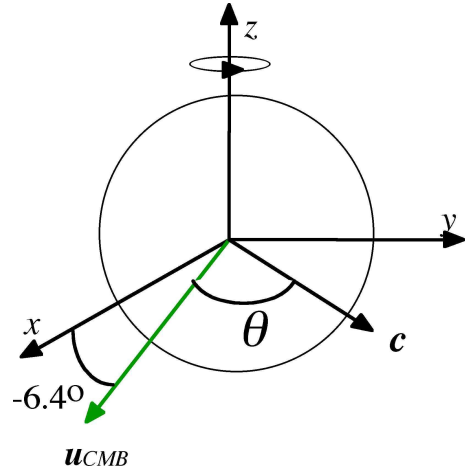}
\caption{This frame is an Earth centred frame in which the spin axis of the earth is the $z$-axis, and the velocity of the Sun with respect to the CMB is defined to have no component in the $y$ direction. Thus, the Earth is spinning at the sidereal rate within this frame, and the angle $\theta$ is shown pictorially, but in general is a function of position and time as the Earth spins and the experiment rotates.}
\label{Earth}
\end{center}
\end{figure}

To calculate the relevant time dependent expressions for $\bf{v}$, velocities are transformed to a geocentric non-rotating (with respect to distant stars) reference frame (denoted as the MM-frame) centred at the centre of mass of the Earth with its $z$-axis perpendicular to the equator, pointing north, the x-axis pointing towards 11.2h right ascension (aligned with the equatorial projection of ${\bf u}$ defined below). A pictorial representation of the frame is shown in Fig. \ref{Earth}.
Classical (Galilean) transformations for the velocities and ${\rm sin}^2\theta$ are sufficient as relativistic terms are of order $\OO2$ and therefore give rise in Eqn. (\ref{cmbvel}) to terms of order $\OO5$. We consider two velocities, the velocity of the sun with respect to the CMB ${\bf  u}$ (declination -6.4$^\circ$, right ascension 11.2 h) and the orbital velocity of the Earth ${\bf v}_o$. Velocities due to the spinning of the Earth and laboratory are much smaller and do not impact on the calculations and may be ignored. Thus, the sum of the two provide the velocity of the laboratory in the universal frame to be inserted in Eqn. (\ref{RMSunit}). In the MM-Earth frame, the CMB velocity is:
\bea
{\bf u}=u\left(\begin{array}{ccc}
			{\rm cos}\phi _\mu \\
			0 \\
			{\rm sin}\phi _\mu
		\end{array}
	 \right)
	 \label{u}
\eea
where $u \approx 377 {\rm km/s}$ and $\phi_\mu \approx -6.4 ^\circ$. To calculate the orbital velocity we first consider the Earth in a barycentric non-rotating frame (BRS) with the $z$-axis perpendicular to the Earth's orbital plane and the $x$-axis pointing towards $0^o$ right ascension (pointing from the Sun to the Earth at the moment of the autumn equinox).
\bea
{\bf v}_o^{BRS}=v_o\left(\begin{array}{ccc}
			-{\rm sin}\lambda_0\\
			{\rm cos}\lambda_0 \\
			0
		\end{array}
	 \right)
\eea
where $v_o \approx 29.78 {\rm km/s}$ is the orbital speed of the Earth, and $\lambda_0=\Omega _\oplus (t-t_o)$ with $\Omega _\oplus \approx 2.0 \ 10^{-7} {\rm rad/s}$ the angular orbital velocity and $t-t_0$ the time since the autumnal equinox. We first transform to a geocentric frame (GRS) that has its $x$-axis aligned with the BRS one but its $z$-axis perpendicular to the equatorial plane of the Earth
\bea
{\bf v}_o^{GRS}=\left(\begin{array}{ccc}
			{v^x}_o^{BRS} \\
			{v^y}_o^{BRS}{\rm cos}\varepsilon - {v^z}_o^{BRS}{\rm sin}\varepsilon  \\
			{v^y}_o^{BRS}{\rm sin}\varepsilon + {v^z}_o^{BRS}{\rm cos}\varepsilon
		\end{array}
	 \right)
\eea
where $\varepsilon \approx 23.27^o$ is the angle between the equatorial and orbital planes of the Earth. We then transform to the MM-Earth frame:
\bea
{\bf v}_o =\left(\begin{array}{ccc}
			{v^x}_o^{GRS}{\rm cos}\alpha _\mu + {v^y}_o^{GRS}{\rm sin}\alpha _\mu \\
			-{v^x}_o^{GRS}{\rm sin}\alpha _\mu + {v^y}_o^{GRS}{\rm cos}\alpha _\mu  \\
			{v^z}_o^{GRS}
		\end{array}
	 \right)
	 \label{vo}
\eea
where $\alpha _\mu \approx 167.9^o$ is the right ascension of ${\bf u}$. Summing the two velocities from Eqns. (\ref{u}) and (\ref{vo}) we obtain the velocity of the lab with respect to the "universe rest frame", transformed to the MM-Earth frame
\bea
{\bf v} =\left(\begin{array}{ccc}
		u{\rm cos}\phi _\mu 
			+ v_o (-{\rm sin}\lambda_0{\rm cos}\alpha _\mu 
			+ {\rm cos}\lambda_0{\rm sin}\alpha _\mu {\rm cos}\varepsilon)\\
			 v_o ({\rm sin}\lambda_0{\rm sin}\alpha _\mu 
			+ {\rm cos}\lambda_0{\rm cos}\alpha _\mu {\rm cos}\varepsilon)\\
			u{\rm sin}\phi _\mu + v_o {\rm cos}\lambda_0{\rm sin}\varepsilon
		\end{array}
	 \right).
	 \label{velocity}
\eea
Substituting in the numeric values gives an orbital velocity of (in $m/s$);
\bea
{\bf v} =\left(\begin{array}{ccc}
		374651 + 5735\cos (\lambda_0) + 29118\sin (\lambda_0)\\
			 -26750\cos (\lambda_0) + 6242\sin (\lambda_0)\\
			-42024 + 11765\cos (\lambda_0)
		\end{array}
	 \right).
	 \label{numvelocity}
\eea

The last calculation to make is the time dependence of the the unit vector $\hat{\bf{l}}$ along the direction of light propagation, which will depend on the configuration of the experiment, including the type of resonator and whether it is rotating in the laboratory or not. In section \ref{RMSExp} we calculate this dependence for a specific experiment, which uses WG modes rotating in the laboratory.

\subsection{Standard Model Extension}
\label{SME}
The SME \cite{Kosto1} conglomerates all possible 
Lorentz-Violating terms and incorporates them in a framework, which is 
an extension of the Standard Model of Particle Physics. There are numerous Lorentz-violating terms per particle sector (i.e. fermions, bosons and photons). However in this work we 
are restricted to the so called minimal ``photon-sector'', which only includes 19 terms. The SME adds additional terms to the Lagrangian of the Standard Model for photons. Where as the standard 
Lagrangian was simply:
\begin{equation}
{\cal L} = - \frac{1}{4} F^{\mu\nu} F_{\mu\nu}
\end{equation}
Under the SME, it becomes \cite{KM}:
\begin{equation}
\label{complicated}
	{\cal L} = 
	- \frac{1}{4} F^{\mu \nu} F_{\mu \nu}
	- \frac{1}{4} (k_F)_{\kappa \lambda \mu \nu} F^{\kappa 
\lambda}F^{\mu \nu}
	+ \frac{1}{2} (k_{AF})^{\kappa} \epsilon_{\kappa \lambda \mu \nu} 
A^\lambda F^{\mu \nu}
\end{equation}
where $A^\lambda$ is the 4-potential. The $(k_{AF})^{\kappa}$ terms 
have the dimensions of mass, and are the CPT odd terms \cite{cptodd}. 
It is argued in \cite{Kosto1} that these should be zero because they 
induce instabilities as they are non-negative in the Lagrangian. There are also astronomical 
measurements \cite{KM} which place stringent limits on 
${k_{AF}}$. From here on these terms are set to zero.

On the other hand, the ${(k_F)}_{\kappa \lambda \mu \nu}$ terms are 
CPT even, dimensionless and have 19 independent terms out of the 256 
possible combinations of $\kappa$, $\lambda$, $\mu$ and $\nu$. Out of 
these independent Lorentz violating terms, 10 combinations have been 
analysed using astrophysical polarisation tests and have an 
upper-limit of $2\times 10^{-32}$ \cite{KM}. This limit is many orders of 
magnitude less than what is expected from laboratory experiments, so these 
terms are set to zero to simplify the calculations and to remain 
consistent with previous results.

We can derive the equations of motion for this system by minimising 
the action given by (\ref{complicated}), using variational techniques 
and the definition $F^{\mu \nu} \equiv \partial_\mu A_\nu - \partial_\nu A_\mu$ and $A_\mu \equiv (\phi, \mathbf{A})$. These 
equations are similar to those of a Maxwellian model in anisotropic 
media instead of a vacuum. In order to express these in a convenient 
form, we form linear combinations of the CPT even term. These are 
given below \cite{KM}:

\begin{eqnarray}
\label{kappadef}
{\kappa_{DE}}^{jk} = -2{(k_F)}^{0j0k};
{\kappa_{HB}}^{jk} = \frac{1}{2} 
\epsilon^{jpq}\epsilon^{krs}{(k_F)}^{pqrs}; \nonumber \\
{\kappa_{DB}}^{jk} = - {\kappa_{HE}}^{kj} = 
{(k_F)}^{0jpq}\epsilon^{kpq}.
\end{eqnarray}

The dynamics of the model can be described in terms of equivalent 
$\mathbf{B}$, $\mathbf{E}$, $\mathbf{H}$ and 
$\mathbf{D}$ fields \cite{KM,WolfGRG} in a vacuum using the 
matrices in (\ref{kappadef}):

\begin{equation}
\label{sixdimmatrixvac}
\left(\begin{array}{c}\mathbf{D}\\\mathbf{H}\end{array}\right) =
\left(\begin{array}{cc}
\epsilon_0 (1+\kappa_{DE}) & \sqrt{\frac{\epsilon_0}{\mu_0}} 
\kappa_{DB}\\
\sqrt{\frac{\epsilon_0}{\mu_0}} \kappa_{HE} & \mu_0^{-1} 
(1+\kappa_{HB})
\end{array}\right)
\left(\begin{array}{c}\mathbf{E}\\\mathbf{B}\end{array}\right)
\end{equation}

Note that (\ref{sixdimmatrixvac}) is rank 6, as the $\kappa$ matrices 
are rank 3 as defined in (\ref{kappadef}). The standard Maxwell 
equations in a vacuum are recovered if these $\kappa$ matrices are 
set to zero.

Thus the effect of the SME in the photon-sector can be interpreted as 
introducing medium-like properties to the vacuum. In the full SME, 
this is considered as an effect from Planck-scale physics in the 
early universe. The $\kappa$ matrices are all position dependent and 
thus act as ``values'' positioned throughout space. If one or more of 
these values are zero, it does not imply the rest are also zero as 
there is no relation between each of the independent components. 
However, there is a linear combination of these components which 
allows us to separate them into birefringent \cite{KM} and 
non-birefringent terms. By eliminating those values which have been 
constrained beyond what we hope to achieve in this experiment, these 
terms can be simply written as in (\ref{ktildes}).

\begin{eqnarray}
\nonumber(\ktilde_{e+})^{jk} &=& \frac{1}{2} {\kappa_{DE}}^{jk} + 
{\kappa_{HB}}^{jk};\, \nonumber(\ktilde_{e-})^{jk} = \frac{1}{2} 
{\kappa_{DE}}^{jk} - {\kappa_{HB}}^{jk} - \frac{1}{3}\delta^{jk} 
(\kappa_{DE})^{ll}\\
\label{ktildes}
(\ktilde_{o+})^{jk} &=& \frac{1}{2} {\kappa_{DB}}^{jk} + 
{\kappa_{HE}}^{jk};\, \ktilde_{o-}^{jk} = \frac{1}{2} 
{\kappa_{DB}}^{jk} - {\kappa_{HE}}^{jk}\\
\nonumber\ktilde_{tr} &=& \frac{1}{3}\delta^{jk} (\kappa_{DE})^{ll} 
\end{eqnarray}

\begin{figure}
\begin{center}
\includegraphics[width=8cm]{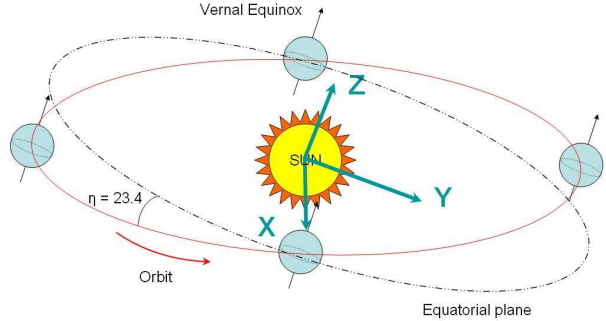}
\caption{The Sun-Centred Celestial Equatorial Frame (SCCEF), with the Earth at different equinoxes during the year. Note that during the vernal equinox, the longitude which is at 
noon has it's $Y$ axis pointing east, while the longitude which is at 
midnight has it's $Y$ axis pointing west, and vice-versa for the autumn equinox.}
\label{SCECF}
\end{center}
\end{figure}

In the above definitions, $\ktilde_{e+}$, $\ktilde_{e-}$ and 
$\ktilde_{tr}$ are parity-even matrices, while $\ktilde_{o+}$ and 
$\ktilde_{o-}$ are the parity-odd matrices. As mentioned in 
\cite{KM,Kost01}, the $\ktilde_{e+}$ and $\ktilde_{o-}$ are constrained 
such that $\left|\ktilde^{JK}\right|\leq2\times10^{-32}$. Thus, both 
$\ktilde_{e+}$ and $\ktilde_{o-}$ are set to zero each time they 
appear in our equations. Also, $\ktilde_{tr}$ is a scalar, which resonator experiments are usually insensitive to \cite{TobarPRD}, and is not considered in this work.

The standard reference frame that we use is the Sun-Centred Celestial Equatorial Frame (SCCEF), which is shown in Fig. \ref{SCECF}. This is the frame in which the sun is at the center, 
and is inertial with respect to the CMB to first order. The axes in 
this frame are labeled $X$, $Y $and $Z$. The $Z$ axis is defined \cite[pg. 6]{KM}\cite[pg. 3]{Bluhm}
to be parallel to the Earth's north pole, or $90^\circ$ declination. 
The $X$ axis points from the sun towards the Earth at the moment of the autumn 
equinox, or $0^\circ$ right ascension (RA) and $0^\circ$ declination, 
while the $Y$ axis is at $90^\circ$ RA and also at $0^\circ$ 
declination, usually taken in the J2000.0 frame.

The convention described in \cite[pg. 18]{KM}, which has the 
raised capital indicies $(J, K)$ in the SCCEF, has been used. Local 
coordinates $x$, $y$ and $z$ are defined on the Earth's surface (at the 
point of the experiment). The $z$ axis is defined as being locally 
normal to the ground, vertically upwards. The $x$ axis points south and 
the $y$ axis points east. These coordinates are denoted by the lowered 
capital indicies $(j, k)$ and they rotate with sidereal period $\Delta 
T_\oplus = \frac{1}{\omega_\oplus} \approx 23$ h $56$ min. There is a 
relation between these two coordinates which is given by the 
following rotation matrix:
\begin{equation}
\label{rotmatrix}
R^{jJ} = 
\left(\begin{array}{ccc}
\cos{\chi} \cos{\omega_\oplus T_\oplus} & \cos{\chi} 
\sin{\omega_\oplus T_\oplus}  & - \sin{\chi} \\
-\sin{\omega_\oplus} T_\oplus & \cos{\omega_\oplus T_\oplus} & 0\\
\sin{\chi} \cos{\omega_\oplus T_\oplus} & \sin{\chi} 
\sin{\omega_\oplus T_\oplus} & \sin{\chi}\\
\end{array}\right)
\end{equation}
Here $\chi$ is the co-latitude of the laboratory from the north pole, and the $T_\oplus$ 
is the time coordinate that is related to the sidereal frequency of 
the Earth. The time $T_\oplus$ is defined in \cite{KM} as any 
time when the $y$ axis and the $Y$ axis align. This has been taken to be 
the first time this occurs after the vernal equinox, which points along the negative $x$-axis.

When searching for leading order violations it is only necessary to consider the spinning of the Earth about itself. However, the orbit of the Earth about the sun may also be 
considered, since it induces Lorentz boosts and we may calculate 
proportional terms to these. Since the Earth moves relatively slowly 
around the Sun compared to the speed of light, the boost terms will 
be suppressed by the velocity with respect to the speed of light 
($\beta_\oplus = \frac{v_\oplus}{c} \approx 10^{-4}$). The boost 
velocity of a point on the Earth's surface is given by the following 
relation:
\begin{equation}
\vec{\beta} = \beta_\oplus \left(\begin{array}{c} \sin{\Omega_\oplus 
T} \\ -\cos{\eta} \cos{\Omega_\oplus T} \\ -\sin{\eta} 
\cos{\Omega_\oplus T} \end{array}\right) +
\beta_L \left(\begin{array}{c} - \sin{\omega_\oplus T} \\ 
\cos{\omega_\oplus T} \\ 0 \end{array}\right)
\end{equation}
Here $\beta_\oplus$ is the value for the boost speed of the orbital motion of the Earth and $\beta_L$ is the boost speed of the lab at the surface of the Earth due to its spin motion. The latter is location dependent, but is less than $1.5\times10^{-6}$ and is zero at the poles ($\eta$ is as defined in 
Fig. \ref{SCECF}). The Lorentz matrix, $\Lambda^\mu_\nu$, that implements the transformation from the SCCEF to the laboratory frame with the sidereal 
rotation $R^{jJ}$ and a boost $\vec{\beta}$ is given by,
\begin{equation}
\Lambda^\mu_\nu =
\left(\begin{array}{cccc}
1 & -\beta^1 					& -\beta^2 & -\beta^3 \\
-(R \cdot {\vec{\beta}})^1 & R^{11} & R^{12} & R^{13} \\
-(R \cdot {\vec{\beta}})^2 & R^{21} & R^{22} & R^{23} \\
-(R \cdot {\vec{\beta}})^3 & R^{31} & R^{32} & R^{33} \\
\end{array}\right)
\end{equation}
After some calculation \cite{KM} the $\kappa$ matrices from the SCCEF (indexed by 
J and K) can be express in terms of the values in the laboratory frame 
(indexed by j and k).
\begin{eqnarray}
\label{transforms}
(\kappa_{DE})^{jk}_{lab} &=& T^{jkJK}_0 (\kappa_{DE})^{JK} - 
T^{kjJK}_1 (\kappa_{DB})^{JK}-T^{jkJK}_1 (\kappa_{DB})^{JK}\\
\label{transformsHB}
(\kappa_{HB})^{jk}_{lab} &=& T^{jkJK}_0 (\kappa_{DE})^{JK} -T^{(kjKJ}_1 (\kappa_{DB})^{JK}-T^{jkKJ}_1 (\kappa_{DB})^{JK}\\
\label{transformsDB}
(\kappa_{DB})^{jk}_{lab} &=& T^{jkJK}_0 (\kappa_{DB})^{JK} + 
T^{kjJK}_1 (\kappa_{DB})^{JK} + T^{jkJK}_1 (\kappa_{HB})^{JK}
\end{eqnarray}
Where:
\begin{eqnarray}
T^{jkJK}_0 &=& R^{jJ} R^{kK}\\
T^{jkJK}_1 &=& R^{jP} R^{kJ} \epsilon^{KPQ} \beta^Q
\end{eqnarray}
Here $\epsilon$ is the standard anti-symmetric tensor. 

In the section \ref{SMEExp} we apply the above to calculate the sensitivity of typical resonator experiment. To do this the sensitivity to the components given in Eqn. (\ref{kappadef}) are derived.

\section{Applying the SME to Resonator Experiments}
\label{SMEExp}

The modified Lagrangian of the SME introduces perturbations of the 
electric and magnetic fields in a vacuum. The unperturbed fields are 
denoted by a zero subscript to distinguish them from the 
Lorentz-violating fields. Putative Lorentz violations are produced by 
motion with respect to a preferred frame, which perturbs the 
fields generating an observable signal. The general 
framework \cite{KM} for denoting the sensitivity of this 
observable signal in the laboratory frame is a linear expression as follows:
\begin{equation}
\label{observable signal}
\delta \mathcal{O} = (\mathcal{M}_{DE})_{lab}^{jk} 
(\kappa_{DE})_{lab}^{jk} + (\mathcal{M}_{HB})_{lab}^{jk} 
(\kappa_{HB})_{lab}^{jk} + (\mathcal{M}_{DB})_{lab}^{jk} 
(\kappa_{DB})_{lab}^{jk}
\end{equation}
The summation over the indicies is implied, and the components of the
$\mathcal{M}_{lab}^{jk}$ matricies are in general a function of time. The observable is dependent on the type of experiment, and in the case of a resonant cavity experiments it is the resonance frequency deviation, $\frac{\delta\nu}{\nu_0}$. Since the laboratory frame and the resonator frame do not necessarily coincide, we first consider $\mathcal{M}_{res}^{jk}$ coefficients in the resonator frame and later relate it to the laboratory and sun-centred frame.

In general, resonators may be constructed from dielectric and magnetic materials. To calculate the
$\mathcal{M}_{res}^{jk}$ matricies for such structures a more general form of 
(\ref{sixdimmatrixvac}) must be considered, which includes the properties of the medium, $\mu_r$ (permeability) and $\epsilon_r$ (permittivity), which are in general second order tensors.
\begin{equation}
\label{preturbedmatrix}
\left(\begin{array}{c}\mathbf{D}\\\mathbf{H}\end{array}\right) =
\left(\begin{array}{cc}
\epsilon_0 (\epsilon_r+\kappa_{DE}) & \sqrt{\frac{\epsilon_0}{\mu_0}} 
\kappa_{DB}\\
\sqrt{\frac{\epsilon_0}{\mu_0}} \kappa_{HE} & \mu_0^{-1} 
(\mu_r^{-1}+\kappa_{HB})
\end{array}\right)
\left(\begin{array}{c}\mathbf{E}\\\mathbf{B}\end{array}\right)
\end{equation}
Here as was derived in \cite{KM}, we assume that 
the fractional frequency shift due to Lorentz violations is given by:
\begin{eqnarray}
\label{freqshift}
\frac{\Delta\nu}{\nu_0} = - \frac{1}{4 \left\langle U \right\rangle} \times \ \ \ \ \ \ \ \ \ \ \ \ \ \ \ \ \ \ \ \ \ \ \ \ \ \ \ \ \ \ \ \ \ \ \ \ \ \ \ \ \ \ \ \ \ \ \ \ \ \ \\
\int_{V} d^3x \left(\epsilon_0 {\mathbf{E}_0^\ast} 
\cdot\kappa_{DE}\cdot{\mathbf{E}_0} - \mu_0^{-1} {\mathbf{B}_0^\ast} 
\cdot\kappa_{HB}\cdot{\mathbf{B}_0} +  2 
Re\left(\sqrt{\frac{\epsilon_0}{\mu_0}} \mathbf{E}_0^\ast \cdot 
\kappa_{DB} \cdot \mathbf{B}_0 \right)\right) \nonumber
\end{eqnarray}
Here $\left\langle U \right\rangle$ is the energy stored in the 
field and is given by the standard electrodynamic integral.
\begin{equation}
\label{freq energy}
\left\langle U \right\rangle = \frac{1}{4} \int_{V} d^3x 
(\mathbf{E}_0 \cdot \mathbf{D}_0^\ast + \mathbf{B}_0 \cdot 
\mathbf{H}_0^\ast)
\end{equation}
In Maxwellian electrodynamics the balance of magnetic and electrical energy in a resonator is equal, so the following is true:
\begin{equation}
\label{equivalent energy}
\left\langle U \right\rangle = \frac{1}{2} \int_{V} d^3x 
\mathbf{E}_0 \cdot \mathbf{D}_0^\ast = 
\frac{1}{2} \int_{V} d^3x \mathbf{B}_0 \cdot \mathbf{H}_0^\ast
\end{equation}
This reduces $\left\langle U \right\rangle$ to an effective 
normalisation factor for either an electric or magnetic filling factor. Also, since the $\kappa$ terms from the integral of (\ref{freqshift}) are only time dependent rather than spatially dependent, the $\kappa$ terms can be removed from the integral. Thus, the final term in (\ref{freqshift}) will be zero since the electric and magnetic fields are orthogonal in a resonant structure. By equating (\ref{observable signal}) and (\ref{freqshift}) in the resonator frame the $(\mathcal{M}_{DB})_{res}^{jk} $ coefficients are calculated to be zero, eliminating the possibility of making a measurement of $\kappa_{tr}$\cite{TobarPRD,Bail}. Assuming the resonator permeability and permittivity have no off-diagonal coefficients (i.e. non-gyrotropic) such that;
\begin{equation}
\label{relative-epsilon}
\epsilon_{r} = \epsilon_0
\left(\begin{array}{ccc}
\epsilon_{x} & 0 & 0 \\
0 & \epsilon_{y} & 0 \\
0 & 0 &  \epsilon_{z}
\end{array}\right) \
\mu_{r} = \mu_0
\left(\begin{array}{ccc}
\mu_{x} & 0 & 0 \\
0 & \mu_{y} & 0 \\
0 & 0 &  \mu_{z}
\end{array}\right)
\end{equation}
the only non-zero coefficients may then be calculated to be
\begin{equation}
\label{scriptm DE def}
(\mathcal{M}_{DE})_{res}^{jj} = - \frac{1}{\epsilon_j} \frac {\int_V 
d^3x \left|{E_0^j}\right|^2} { 2 \int_V d^3x 
{\mathbf{E}_0}^\ast \cdot \mathbf{E}_0} = - \frac{Pe_j}{2 
\epsilon_{j}}
\end{equation}
\begin{equation}
\label{scriptm HB def}
(\mathcal{M}_{HB})_{res}^{jj} = \mu_j \frac {\int_V 
d^3x \left|{H_0^j}\right|^2} { 2 \int_V d^3x 
{\mathbf{H}_0}^\ast \cdot \mathbf{H}_0} =\mu_j \frac{Pm_j}{2}
\end{equation}
Thus the $\mathcal{M}_{DE}$ and $\mathcal{M}_{HB}$ matricies are diagonal 
and simply related to the electric and magnetic energy filling factors, $Pe_j$ and $Pm_j$ respectively \cite{WolfGRG}. 
In general a resonator may consist of more than one material, and may include vacuum. In this case (\ref{scriptm DE def}) and (\ref{scriptm HB def}) may be written more generally ($s$ is the number of different materials including vacuum).
\begin{eqnarray}
\label{scriptmDE}
(\mathcal{M}_{DE})_{res}^{jj} & = & - \sum_{i=1}^s \frac{Pe^i_j}{2 
\epsilon^i_j} \\
\label{scriptmHB}
(\mathcal{M}_{HB})_{res}^{jj} & = & \sum_{i=1}^s \frac {\mu^i_j Pm^i_j}{2}
\end{eqnarray}

To measure the resonant frequency it is necessary excite electromagnetic fields inside the resonator and then compare it against a similar frequency. To be sensitive to violations of LLI, the comparison frequency must be generated by a source which exhibits a different dependence on Lorentz violations in the photon sector. For example, an atomic standard (such as a hydrogen maser) may operate in a mode which is not sensitive to Lorentz violations \cite{WolfGRG,Wolf04}. Alternatively, the resonant frequency may be compared against another resonator designed to have a different dependence. The latter can be achieved by orientating two identical resonators orthogonally \cite{Muller}, or by exciting two modes in a matter filled resonator with orthogonal polarizations. In both cases the field components must be considered with respect to the laboratory frame and not the resonator. For such an experiment the observable becomes the frequency difference (between a resonator labeled $a$ and $b$) such that;
\begin{equation}
\label{beat observable}
\delta \mathcal{O} = \frac{\delta \nu_a}{\nu_a} - \frac{\delta 
\nu_b}{\nu_b}
\end{equation}
Thus, with respect to the laboratory frame, the effective $(\mathcal{M}_{DE})_{lab}$ and $(\mathcal{M}_{HB})_{lab}$ matricies consistent with (\ref{observable signal}) become:
\begin{eqnarray}
\label{scriptmde}
(\mathcal{M}_{DE})_{a-b} = \ \ \ \ \ \ \ \ \ \ \ \ \ \ \ \ \ \ \ \ \ \ \ \ \ \ \ \ \ \ \ \ \ \ \ \ \ \ \ \ \ \ \ \\
\left(\begin{array}{ccc}
(\mathcal{M}_{DE})_a^{xx} - (\mathcal{M}_{DE})_b^{xx} & 0 & 0 \\
0 & (\mathcal{M}_{DE})_{a}^{yy} - (\mathcal{M}_{DE})_b^{yy} & 0  \\
0 & 0 &  (\mathcal{M}_{DE})_{a}^{zz} - (\mathcal{M}_{DE})_b^{zz} \nonumber 
\end{array}\right)
\\ \nonumber 
\end{eqnarray}
\begin{eqnarray}
\label{scriptmhb}
(\mathcal{M}_{HB})_{a-b} = \ \ \ \ \ \ \ \ \ \ \ \ \ \ \ \ \ \ \ \ \ \ \ \ \ \ \ \ \ \ \ \ \ \ \ \ \ \ \ \ \ \ \ \\
\left(\begin{array}{ccc}
(\mathcal{M}_{HB})_a^{xx} - (\mathcal{M}_{HB})_b^{xx} & 0 & 0 \\
0 & (\mathcal{M}_{HB})_{a}^{yy} - (\mathcal{M}_{HB})_b^{yy} & 0  \\
0 & 0 &  (\mathcal{M}_{HB})_{a}^{zz} - (\mathcal{M}_{HB})_b^{zz} \nonumber 
\end{array}\right)
\\ \nonumber 
\end{eqnarray}
These equations are general for any resonator experiments, including Fabry-Perot and microwave cavity experiments, and simplify the analysis for complex resonator configurations, such as whispering gallery mode resonators. Only the electric and magnetic filling factors need to be calculated to determine the sensitivity coefficients to the observable, which is possible using standard numerical techniques \cite{krupka}. 

To determine the sensitivity of stationary laboratory experiments one calculates the time dependence of (\ref{beat observable}) due to the sidereal and orbital motion of the Earth around the Sun in terms of the Sun-centred coefficients given in (\ref{ktildes}) and (\ref{transforms}). This calculation has already been done in \cite{KM,WolfGRG,Muller2} and will not be repeated here. In the following subsection we generalize this analysis to rotating experiments.

\subsection{Rotation in the Laboratory Frame}

Non-rotating experiments \cite{Lipa,Muller,Wolf04} that rely on Earth rotation to modulate a Lorentz violating effect are not optimal for two reasons. Firstly, the sensitivity is proportional to the noise in the system at the modulation frequency, typically best for microwave resonator-oscillators and Fabry-Perot stabilized lasers for periods between 10 to 100 seconds. Secondly, the sensitivity is
proportional to the square root of the number of periods of the modulation signal, therefore
taking a relatively long time to acquire sufficient data. Thus, by rotating the experiment the data integration rate is increased and the relevant signals are translated to the optimal operating regime \cite{Mike}. For rotation in the laboratory frame the $(\mathcal{M})_{lab}^{jk}$ coefficients become a function of time and depend on the axis of rotation. In the laboratory it is most practical to rotate around the axis of the gravitational field to reduce gravity induced perturbation of the experiment. Thus, our analysis includes rotation about the laboratory $z$-axis. If we set the time, $t = 0$, to be defined when the experiment and laboratory axes are aligned, and we only consider the time varying components (i.e. the most sensitive ones induced by rotation), then for clock-wise rotation of $\omega_R$ rads/sec, 
(\ref{scriptmde}) and (\ref{scriptmhb})become:
\begin{eqnarray}
(\mathcal{M}_{DE})_{lab} & = & 
\left(\begin{array}{ccc}
\mathcal{S}_{DE} \cos(2 \omega_R t) & -\mathcal{S}_{DE} \sin(2 
\omega_R t) & 0\\
-\mathcal{S}_{DE} \sin(2 \omega_R t) & -\mathcal{S}_{DE} \cos(2 
\omega_R t) & 0\\
0 & 0 & 0
\end{array}\right)\\
\label{scriptmDE lab}
\mathcal{S}_{DE}  & = & \frac{1}{2}\left((\mathcal{M}_{DE})_a^{xx}-(\mathcal{M}_{DE})_a^{yy}-(\mathcal{M}_{DE})_b^{xx}+(\mathcal{M}_{DE})_b^{yy}\right)\\
(\mathcal{M}_{HB})_{lab} & = & 
\left(\begin{array}{ccc}
\mathcal{S}_{HB} \cos(2 \omega_R t) & -\mathcal{S}_{HB} \sin(2 
\omega_R t) & 0\\
-\mathcal{S}_{HB} \sin(2 \omega_R t) & -\mathcal{S}_{HB} \cos(2 
\omega_R t) & 0\\
0 & 0 & 0
\end{array}\right)\\
\mathcal{S}_{HB}  & = & \frac{1}{2}\left((\mathcal{M}_{HB})_a^{xx}-(\mathcal{M}_{HB})_a^{yy}-(\mathcal{M}_{HB})_b^{xx}+(\mathcal{M}_{HB})_b^{yy}\right)
\label{scriptmHB lab}
\end{eqnarray}
Note that if one resonator is tested with respect to a stationary generated frequency, then the $(\mathcal{M})_i^{jj}$ coefficients in the definition of $\mathcal{S}_{HB}$ and $\mathcal{S}_{HB}$ pertaining to that frequency must be set to zero.

To determine the time dependence of the observable (\ref{beat observable}) we follow the same procedure as presented in subsection \ref{SME} to transform the $\tilde{\kappa}$ matricies given in (\ref{ktildes}) to the $\kappa$ matricies in the laboratory given by (\ref{transforms}), (\ref{transformsHB}) and (\ref{transformsDB}). We then substitute (\ref{scriptmDE lab}) to (\ref{scriptmHB lab}) into (\ref{observable signal}) to calculate the time dependence of the observable of (\ref{beat observable}). This is a tedious process and the details are omitted. Essentially, because the ($\mathcal{M}_{DE})_{lab}$ and ($\mathcal{M}_{HB})_{lab}$ matricies are time dependent at $2\omega_R$, the observable signals are at frequencies close to this value and are summarized in Table \ref{sense-coefficients}. Here the frequency of Earth rotation is defined as $\omega_\oplus$, and orbit around the sun as $\Omega_\oplus$. The $\omega_\oplus$ is commonly referred to as the sidereal frequency, while the $\Omega_\oplus$ is referred to as the annual frequency. We also define the sensitivity factor, $\mathcal{S}$ of the experiment as:
\begin{equation}
\mathcal{S}=\mathcal{S}_{HB} -\mathcal{S}_{DE}
\label{totsens}
\end{equation}

\begin{table}[htbp]
\begin{center}
\begin{tabular}{|c||l|l|}
\hline
$\omega_i$ & Cosine Coefficient $C_{\omega_i}/\mathcal{S}$ & Sine Coefficient 
$S_{\omega_i}/\mathcal{S}$ \\
\hline
\hline
$2\omega_R$ & $\frac{3}{2} \sin^2(\chi) \ktilde_{e-}^{ZZ}$ & $2 \beta_L \sin(\chi) \ktilde_{o+}^{XY}$\\
\hline
$2\omega_R+ \Omega_\oplus$ & $-\frac{1}{2} \beta_\oplus \sin^2(\chi)\times$ & $-\frac{1}{2} 
\beta_\oplus \sin^2(\chi) \ktilde_{o+}^{YZ}$\\
 &$(2\sin(\eta)\ktilde_{o+}^{XY} + \cos(\eta) \ktilde_{o+}^{XZ})$ & \\
 \hline
$2 \omega_R-\Omega_\oplus$ & $-\frac{1}{2} \beta_\oplus \sin^2(\chi)\times $ & $\frac{1}{2} 
\beta_\oplus \sin^2(\chi) \ktilde_{o+}^{YZ}$\\
 &$(2\sin(\eta) \ktilde_{o+}^{XY} + \cos(\eta) \ktilde_{o+}^{XZ})$ & \\
\hline
\hline
$2 \omega_R+\omega_\oplus$ & $-2 \sin^2(\frac{\chi}{2})\times$ & $-2 \sin^2(\frac{\chi}{2}) 
(\sin(\chi) \ktilde_{e-}^{YZ} - \beta_L \ktilde_{o+}^{YZ})$\\
 &$ (-\beta_L \ktilde_{o+}^{XZ} + \sin(\chi) \ktilde_{e-}^{XZ})$ & \\
 \hline
$2\omega_R+\omega_\oplus+\Omega_\oplus$ & $-2 \beta_\oplus \cos(\frac{\chi}{2}) \sin(\eta) 
\sin^3(\frac{\chi}{2}) \ktilde_{o+}^{YZ}$ & $4 \beta_\oplus 
\cos(\frac{\chi}{2}) \sin(\frac{\eta}{2}) \sin^3(\frac{\chi}{2})\times$\\
 & &$(\sin(\frac{\eta}{2}) \ktilde_{o+}^{XY} + \cos(\frac{\eta}{2})\ktilde_{o+}^{XZ})$ \\
 \hline
$2 \omega_R + \omega_\oplus- \Omega_\oplus$ & $-2 \beta_\oplus \cos(\frac{\chi}{2}) \sin(\eta) 
\sin^3(\frac{\chi}{2}) \ktilde_{o+}^{YZ}$ & $-4 \beta_\oplus 
\cos(\frac{\chi}{2}) \cos(\frac{\eta}{2}) \sin^3(\frac{\chi}{2})\times$\\
 & &$(\cos(\frac{\eta}{2}) \ktilde_{o+}^{XY} - \sin(\frac{\eta}{2})\ktilde_{o+}^{XZ})$ \\
\hline
\hline
$2 \omega_R+2\omega_\oplus$ & $-\sin^4(\frac{\chi}{2}) (\ktilde_{e-}^{XX} - 
\ktilde_{e-}^{YY})$ & $-2 \sin^4(\frac{\chi}{2}) \ktilde_{e-}^{XY}$\\
\hline
$2\omega_R+2\omega_\oplus+\Omega_\oplus$ & $2 \beta_\oplus \sin^2(\frac{\eta}{2}) 
\sin^4(\frac{\chi}{2}) \ktilde_{o+}^{XZ}$ & $2 \beta_\oplus 
\sin^2(\frac{\eta}{2}) \sin^4(\frac{\chi}{2}) \ktilde_{o+}^{YZ}$\\
\hline
$2 \omega_R + 2\omega_\oplus- \Omega_\oplus$ & $-2 \beta_\oplus \cos^2(\frac{\eta}{2}) 
\sin^4(\frac{\chi}{2}) \ktilde_{o+}^{XZ}$ & $-2 \beta_\oplus 
\cos^2(\frac{\eta}{2}) \sin^4(\frac{\chi}{2}) \ktilde_{o+}^{YZ}$\\
\hline
\hline
$2 \omega_R-\omega_\oplus$ & $2 \cos^2(\frac{\chi}{2}) (\beta_L 
\ktilde_{o+}^{XZ} + \sin(\chi) \ktilde_{e-}^{XZ})$ & $-2\cos^2(\frac{\chi}{2}) (\beta_L \ktilde_{o+}^{YZ} + \sin(\chi) \ktilde_{e-}^{YZ})$\\
\hline
$2\omega_R-\omega_\oplus+\Omega_\oplus$ & $2 \beta_\oplus \cos^3(\frac{\chi}{2}) \sin(\eta) 
\sin(\frac{\chi}{2}) \ktilde_{o+}^{YZ}$ & $-4 \beta_\oplus 
\cos(\frac{\eta}{2}) \cos^3(\frac{\chi}{2}) \sin(\frac{\chi}{2})\times$\\
 & &$(\cos(\frac{\eta}{2}) \ktilde_{o+}^{XY} - \sin(\frac{\eta}{2})\ktilde_{o+}^{XZ})$ \\
\hline
$2 \omega_R - \omega_\oplus- \Omega_\oplus$ & $2 \beta_\oplus \cos^3(\frac{\chi}{2}) \sin(\eta) 
\sin(\frac{\chi}{2}) \ktilde_{o+}^{YZ}$ & $4 \beta_\oplus 
\sin(\frac{\eta}{2}) \cos^3(\frac{\chi}{2}) \sin(\frac{\chi}{2})\times$\\
 & &$(\sin(\frac{\eta}{2}) \ktilde_{o+}^{XY} + \cos(\frac{\eta}{2})\ktilde_{o+}^{XZ})$ \\
\hline
\hline
$2 \omega_R-2\omega_\oplus$ & $-\cos^4(\frac{\chi}{2}) (\ktilde_{e-}^{XX} - 
\ktilde_{e-}^{22})$ & $2 cos^4(\frac{\chi}{2}) \ktilde_{e-}^{XY}$\\
\hline
$2\omega_R-2\omega_\oplus+\Omega_\oplus$ & $-2 \beta_\oplus \cos^2(\frac{\eta}{2}) 
\cos^4(\frac{\chi}{2}) \ktilde_{o+}^{XZ}$ & $2 \beta_\oplus 
\cos^2(\frac{\eta}{2}) \cos^4(\frac{\chi}{2}) \ktilde_{o+}^{YZ}$\\
\hline
$2 \omega_R - 2\omega_\oplus- \Omega_\oplus$ & $2 \beta_\oplus \sin^2(\frac{\eta}{2}) 
\cos^4(\frac{\chi}{2}) \ktilde_{o+}^{XZ}$ & $-2 \beta_\oplus 
\sin^2(\frac{\eta}{2}) \cos^4(\frac{\chi}{2}) \ktilde_{o+}^{YZ}$\\
\hline
\end{tabular}
\end{center}
\caption{Normalized sensitivities with respect to the experiment sensitivity factor $\mathcal{S}$ for all predicted frequency modulated components.}
\label{sense-coefficients}
\end{table}

\begin{figure}[ht]
\begin{center}
\includegraphics[width=8cm]{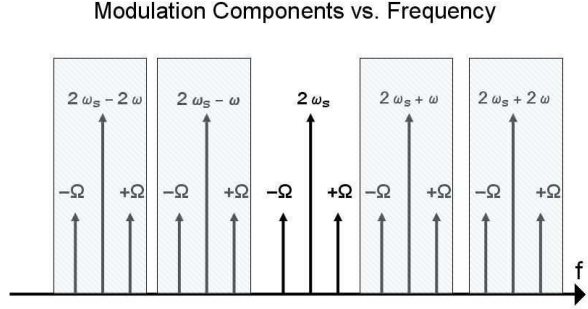}
\caption{This ``frequency stick diagram'' shows a schematic of the frequency 
modulation components of the beat frequency in a convenient form. The 
sidebands offset by $\pm \Omega_\oplus$ have been trimmed for brevity. Each frequency has two degrees of freedom, 
i.e. a sine and a cosine term, with the phase set by a time $t = 0$ set with respect to the SCCEF.}
\label{frequency stick}
\end{center}
\end{figure}

To decorrelate all side bands, more than one year of data is necessary. In this case we have eight unknown $\tilde{\kappa}$ coefficients and thirty possible individual measurements listed in Table \ref{sense-coefficients}, which is an over parameterization. For short data sets (less than a year) we do not have enough information to satisfy the Nyquist condition to distinguish between frequencies that differ by the annual offset (collected in the same blocks). Thus, to make a short data set approximation, we collect the sidebands together (see Fig. \ref{frequency stick}).
The short data set approximation is achieved by knowing the angle of the orbit, $\Phi = \Omega_\oplus t$, in the sun-centered frame with respect to the negative $X$-axis (which occurs at the vernal equinox as shown in Fig. \ref{SCECF}), and then taking a Taylor series expansion around that angle. Here we define the phase of the combined rotational and 
sidereal term as $\theta$ and $\Phi_0$ as the value of 
$\Phi$ when a short data set is taken. Since $\delta \Phi 
\equiv \Phi - \Phi_0$ is small with respect to $2\pi$, via the double angle rule we can derive the following relationships:
\begin{eqnarray*}
\sin(\theta \pm (\delta \Phi + \Phi_0))
& = & \sin(\theta \pm \delta \Phi) \cos(\Phi_0) \pm \cos(\theta \pm 
\delta \Phi) \sin(\Phi_0)\\
\ &\approx & \sin(\theta) \cos(\Phi_0) \pm \cos(\theta) \sin(\Phi_0)
\\
\cos(\theta \pm (\delta \Phi + \Phi_0))
& = & \cos(\theta \pm \delta \Phi) \cos(\Phi_0) \mp \sin(\theta \pm 
\delta \Phi) \sin(\Phi_0)\\
\ & \approx & \cos(\theta) \cos(\Phi_0) \mp \sin(\theta) \sin(\Phi_0)
\\
\end{eqnarray*}
Now we can combine the sidebands as shown in Fig. \ref{frequency stick} by applying the above relationships to eliminate the dependence on $\Omega_\oplus$. In this case the components from Table \ref{sense-coefficients} decompose to those listed in Table \ref{combined}.
\begin{table}
\begin{center}
\begin{tabular}{|l||c|}
\hline
 & Coefficient (normalized with respect to $\mathcal{S}$)\\
\hline
\hline
$S_{2 \omega_R - 2 \omega_\oplus}$ & $- \cot^4(\frac{\chi}{2})\times S_{2 \omega_R + 2 \omega_\oplus}$\\
\hline
$C_{2 \omega_R - 2 \omega_\oplus}$ & $\cot^4(\frac{\chi}{2})\times C_{2 \omega_R + 2 \omega_\oplus}$\\
\hline
$S_{2 \omega_R - \omega_\oplus}$ & $-2 \cos^2(\frac{\chi}{2})\times$\\
 & $( \beta_L \ktilde_{o+}^{YZ} + \sin(\chi) (\beta_\oplus \cos(\Phi_0) (\cos(\eta) \ktilde_{o+}^{XY} - \sin(\eta) \ktilde_{o+}^{XZ})+\ktilde_{e-}^{YZ}))$\\
\hline
$C_{2 \omega_R - \omega_\oplus}$ & $2 \cos^2(\frac{\chi}{2})\times $ \\
 &$(\beta_L \ktilde_{o+}^{XZ} + \sin(\chi) ( - \beta_\oplus \sin(\Phi_0) \ktilde_{o+}^{XY} + \beta_\oplus \cos(\Phi_0) \sin(\eta) \ktilde_{o+}^{YZ} + \ktilde_{e-}^{XZ}))$\\
\hline
$S_{2 \omega_R}$ & $2 \beta_L \sin(\chi) \ktilde_{o+}^{XY}$\\
\hline
$C_{2 \omega_R}$ & -$\frac{1}{2} \sin^2(\chi)(2 \beta_\oplus \cos(\Phi_0) (2 \sin(\eta) \ktilde_{o+}^{XY} +\cos(\eta) \ktilde_{o+}^{XZ})+2 \beta_\oplus \sin(\Phi_0) \ktilde_{o+}^{YZ} ) $\\
  & $+\frac{3}{2}  \sin^2(\chi)\ktilde_{e-}^{ZZ})$\\
\hline
$S_{2 \omega_R + \omega_\oplus}$ & $-2 \sin^2(\frac{\chi}{2})\times$\\
 & $(-\beta_L \ktilde_{o+}^{YZ} + \sin(\chi) (\beta_\oplus \cos(\Phi_0)(\cos(\eta) \ktilde_{o+}^{XY} - \sin(\eta) \ktilde_{o+}^{XZ}) + \ktilde_{e-}^{YZ}))$\\
\hline
$C_{2 \omega_R + \omega_\oplus}$ & $-2 \sin^2(\frac{\chi}{2})\times$\\
 &$(-\beta_L \ktilde_{o+}^{XZ} + \sin(\chi)(-\beta_\oplus \sin(\Phi_0) \ktilde_{o+}^{XY} + \beta_\oplus \cos(\Phi_0) \sin(\eta) \ktilde_{o+}^{YZ} + \ktilde_{e-}^{XZ}))$\\
\hline
$S_{2 \omega_R + 2 \omega_\oplus}$ & $-2 \sin^4(\frac{\chi}{2}) (\beta_\oplus \sin(\Phi_0) \ktilde_{o+}^{XZ} + \beta_\oplus \cos(\eta) \cos(\Phi_0) \ktilde_{o+}^{YZ} + \ktilde_{e-}^{XY})$\\
\hline
$C_{2 \omega_R + 2 \omega_\oplus}$ & $-\sin^4(\frac{\chi}{2}) (2 \beta_\oplus \cos(\eta) \cos(\Phi_0) \ktilde_{o+}^{XZ} - 2 \beta_\oplus \sin(\Phi_0) \ktilde_{o+}^{YZ} + (\ktilde_{e-}^{XX} - \ktilde_{e-}^{YY})) $\\
\hline
\end{tabular}
\end{center}
\caption{Normalized sensitivities with respect to the experiment sensitivity factor $\mathcal{S}$ for all predicted frequency modulated components using the short data set approximation.}
\label{combined}
\end{table}
The first feature to notice is that the $2 \omega_R\pm 2 \omega_\oplus$ sidebands are redundant. One might also expect the $2 \omega_R\pm \omega_\oplus$ sidebands to be redundant as well. The only reason they are not is because we have taken into account the velocity of the laboratory due to the Earth spinning on its axis, $\beta_L$. In fact it turns out that it is not useful to keep this term because $\beta_\oplus$ is two orders of magnitude larger and when one applies the data analysis procedures the sensitivities will be degraded if the analysis depends on the $\beta_L$ terms for the uniqueness of the solution. Thus, since it makes no practical sense to keep these terms, we set them to zero. For this case the coefficients are listed in Table \ref{approx}.
\begin{table}
\begin{center}
\begin{tabular}{|l||c|}
\hline
 & Coefficient (normalized with respect to $\mathcal{S}$)\\
\hline
\hline
$S_{2 \omega_R - 2 \omega_\oplus}$ & $- \cot^4(\frac{\chi}{2})\times S_{2 \omega_R + 2 \omega_\oplus}$\\
\hline
$C_{2 \omega_R - 2 \omega_\oplus}$ & $\cot^4(\frac{\chi}{2})\times C_{2 \omega_R + 2 \omega_\oplus}$\\
\hline
$S_{2 \omega_R - \omega_\oplus}$ & $\cot^2(\frac{\chi}{2})\times S_{2 \omega_R +  \omega_\oplus}$\\
\hline
$C_{2 \omega_R - \omega_\oplus}$ & $-\cot^2(\frac{\chi}{2})\times C_{2 \omega_R +  \omega_\oplus}$ \\
\hline
$S_{2 \omega_R}$ & ----  \\
\hline
$C_{2 \omega_R}$ & $-\frac{1}{2} \sin^2(\chi)(2 \beta_\oplus \cos(\Phi_0) (2 \sin(\eta) \ktilde_{o+}^{XY} +\cos(\eta) \ktilde_{o+}^{XZ})+2 \beta_\oplus \sin(\Phi_0) \ktilde_{o+}^{YZ} ) $\\
  & $+\frac{3}{2}  \sin^2(\chi)\ktilde_{e-}^{ZZ})$\\
\hline
$S_{2 \omega_R + \omega_\oplus}$ & $-2 \sin^2(\frac{\chi}{2}) ( \sin(\chi) (\beta_\oplus \cos(\Phi_0)(\cos(\eta) \ktilde_{o+}^{XY} - \sin(\eta) \ktilde_{o+}^{XZ}) + \ktilde_{e-}^{YZ}))$\\
\hline
$C_{2 \omega_R + \omega_\oplus}$ & $-2 \sin^2(\frac{\chi}{2})(\sin(\chi)(-\beta_\oplus \sin(\Phi_0) \ktilde_{o+}^{XY} + \beta_\oplus \cos(\Phi_0) \sin(\eta) \ktilde_{o+}^{YZ} + \ktilde_{e-}^{XZ}))$\\
\hline
$S_{2 \omega_R + 2 \omega_\oplus}$ & $-2 \sin^4(\frac{\chi}{2}) (\beta_\oplus \sin(\Phi_0) \ktilde_{o+}^{XZ} + \beta_\oplus \cos(\eta) \cos(\Phi_0) \ktilde_{o+}^{YZ} + \ktilde_{e-}^{XY})$\\
\hline
$C_{2 \omega_R + 2 \omega_\oplus}$ & $-\sin^4(\frac{\chi}{2}) (2 \beta_\oplus \cos(\eta) \cos(\Phi_0) \ktilde_{o+}^{XZ} - 2 \beta_\oplus \sin(\Phi_0) \ktilde_{o+}^{YZ} + (\ktilde_{e-}^{XX} - \ktilde_{e-}^{YY})) $\\
\hline
\end{tabular}
\end{center}
\caption{Normalized sensitivities with respect to the experiment sensitivity factor $\mathcal{S}$ for all predicted frequency modulated components, using the short data set approximation and neglecting components of order $\beta_L$.}
\label{approx}
\end{table}

For data sets of less than one year, the components in Table \ref{approx} may be used to set upper limits on the $\tilde{\kappa}$ coefficients in the SME. Since there are only five possible independent components, to set limits on eight coefficients we use the same technique as adopted by  Lipa et. al. \cite{Lipa}. The $\tilde{\kappa}_{o+}$ boost coefficients are set to zero to calculate limits on the $\tilde{\kappa}_{e-}$ isotropy coefficients and vice versa. This technique assumes no correlation between the isotropy and boost coefficients. It would be unlikely that a cancelation of Lorentz violating effects would occur, as this would necessitate a fortuitous relationship between the coefficients of the same order of value as the boost suppression coefficient (i.e. orbit velocity, $\beta_\oplus$), and consistent with the correct linear combinations as presented in Tab \ref{approx}.

Another practical point is that the largest systematic effect occurs at $2\omega_R$. Thus, when setting the limits on the three $\ktilde_{o+}^{XY}$, $\ktilde_{o+}^{XZ}$ and $\ktilde_{o+}^{YZ}$ coefficients we only use the data collected at $2 \omega_R \pm \omega_\oplus$ and $2 \omega_R \pm 2\omega_\oplus$ frequencies. Likewise for the $\ktilde_{e-}^{XY}$, $\ktilde_{e-}^{XZ}$, $\ktilde_{e-}^{YZ}$ and $(\ktilde_{e-}^{XX} - \ktilde_{e-}^{YY})$ coefficients. These are the same coefficients that have had limits set by the non-rotating experiments \cite{Muller}\cite{Lipa}\cite{Wolf04}. The remaining coefficient $\ktilde_{e-}^{ZZ} (\equiv \ktilde_{e-}^{XX} + \ktilde_{e-}^{YY})$ can only be set amongst a systematic signal at $2 \omega_R$, which is in general much greater than the statistical uncertainties at the other frequencies. In this case we can assume all coefficients are zero except for the $\ktilde_{e-}^{ZZ}$ coefficient. However, it is not straight forward to set a limit on any putative Lorentz violation amongst a large systematic as one can not be sure if the systematic signal actually cancels an effect. Since the signal at $2\omega_R$ is dominated by systematic effects, it is likely that its phase and amplitude will vary across different data sets. In this case the systematic signal from multiple data sets can be treated statistically to place an upper limit on $\ktilde_{e-}^{ZZ}$.  In our experiment we use this technique to set an upper of $2.1(5.7)\times10^{-14}$ \cite{PRL} (see section \ref{UWA}).

\subsection{Phase with respect to the SCCEF}
\label{phSCECF}

To extract the $\kappa$ components of the SME out of our observed
signal we first need to
determine the relevant $C_{2\omega_i}$ and $S_{2\omega_i}$
coefficients listed in Table \ref{sense-coefficients}. This in turn
requires us to know the phase of the experiment's orientation with
respect to the SCCEF. In this section we will derive an expression
for this phase in terms of the time origins of the experiment's
rotation, the Earth's sidereal rotation, and the orbit of the Earth
around the Sun.

In general, we are interested in the frequency components
\begin{equation}
2\omega_{[a,b]} = 2\omega_R + a\omega_{\oplus} + b\Omega_{\oplus}
\end{equation}
where $a$ and $b$ take on values in the domains
\begin{equation}
a \in [-2,2], b \in [-1,1]
\end{equation}
Thus to determine the $C_{2\omega_{[a,b]}}$ coefficient we fit the
data with a model of the form
\begin{equation}
\cos(2\omega_R T_R + a\omega_{\oplus} T_{\oplus} + b\Omega_{\oplus}
T) \label{cos_phase}
\end{equation}
where $T_R$ is the experiment's rotation time, $T_{\oplus}$ is the
sidereal time, and $T$ is the time since the vernal equinox.

To simplify our analysis we aim to transform this expression to the
form $\cos(\alpha t + \phi)$. To achieve this we note that the
difference $\delta_{R}$ between the experiment's rotation time $T_R$
and the time since the vernal equinox $T$ is constant over the
course of the measurement, as determined by the initial
configuration of the experiment, and we may write,
\begin{equation}
\delta_{R} = T_R - T. \label{deltaR}
\end{equation}
Similarly the sidereal time and the time since the vernal equinox
are related by $\delta_{\oplus}$,
\begin{equation}
\delta_{\oplus} = T_{\oplus} - T \label{deltaO}
\end{equation}
By combining (\ref{cos_phase}), (\ref{deltaR}) and (\ref{deltaO}) we
arrive at an expression of the desired form.
\begin{eqnarray*}
\cos(2\omega_R T_R + a\omega_{\oplus} T_{\oplus} + b\Omega_{\oplus}
T) &=&\cos(2\omega_R (\delta_{R}+T) + a\omega_{\oplus}
(\delta_{\oplus} + T) + b\Omega_{\oplus} T)\\ &=&\cos((2\omega_R+a\omega_{\oplus}+b\Omega_{\oplus})T \\ & &  \ \ \ \ \ +2\omega_R \delta_{R}+a\omega_{\oplus} \delta_{\oplus})
\end{eqnarray*}
Thus we can account for the phase of the experiment relative to the
SCCEF by determining $\delta_R$ and $\delta_{\oplus}$. The origin of
the experiment's rotation time $T_R$ is defined to be the instant at
which the axis of symmetry of the first resonator (resonator $a$) is aligned with
the local $y$ axis. Our experiment has been designed such that the
time origin of the data acquisition coincides with the same event,
rendering $\delta_R = 0$ in our case.
\begin{figure}[htbp]
\begin{center}
\includegraphics[width=7cm]{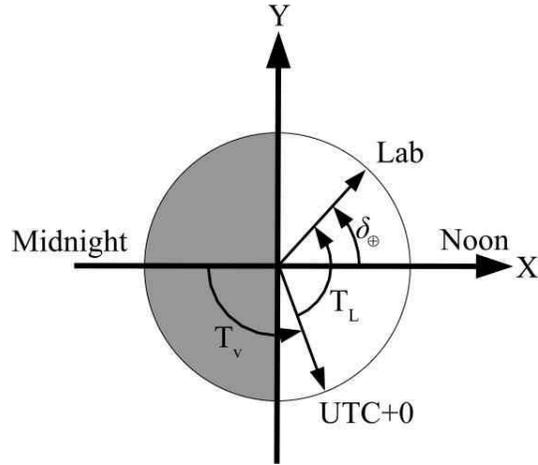}
\caption{Diagram showing meridians and angles.}
\label{SMEphase}
\end{center}
\end{figure}

We also need to obtain $\delta_\oplus$ for the sidereal rotation. We
define $T_\oplus=0$ as in \cite{KM} to be the instant the local $y$
axis and the SCCEF $Y$ axis are aligned (noon) in the laboratory (see
figure \ref{SCECF}). Let us define $T_v$ to be the time in seconds
after midnight UTC+0, at which the vernal equinox has occurred in
the J2000.0 frame \cite{KM}. For convenience we also define our
longitude $T_L$ in terms of sidereal seconds from midnight (in the
case of our laboratory $T_l = 115.826^\circ \times \frac{23 hr 56
min}{360^\circ} = 27721 sec$). There exists a special location whose
meridian is at noon at the vernal equinox. For this special location
(during the vernal equinox), $\delta_\oplus = 0$ since the time when
the $y$ and $Y$ axes align and the vernal equinox are the same. 
We see geometrically that any longitude greater than this meridian will
have positive $\delta_\oplus$, otherwise if the longitude is less
than this meridian it would have negative $\delta_\oplus$. As shown
in Fig. \ref{SMEphase}, we can now derive an expression for
$\delta_{\oplus}$.
\begin{equation}
\delta_{\oplus} = T_L + T_v - \frac{23hr56min}{2}
\end{equation}
Hence we are able to determine the phase of the experiment's
orientation relative to the SCCEF.

\section{Comparison of sensitivity of various resonator experiments in the SME}

In this section we show how the general analysis may be applied to some common resonator configurations for testing LLI. Also, we propose some new configurations based on exciting two modes in matter filled resonators. The comparison is made by calculating the sensitivity parameter $\mathcal{S}$ of the resonator using Eqns. (\ref{scriptmDE}) to (\ref{totsens}). Note that the sign of the  $\mathcal{S}$ factor depends on the definition of the first resonator. Practically this will need to be the resonator that exhibits the largest value of frequency. In this work, where appropriate, we assume the first resonator is aligned along the $y$-axis.

\subsection{Fabry-Perot resonators}
\label{FPr}

\begin{figure}[htbp]
\begin{center}
\includegraphics[width=8cm]{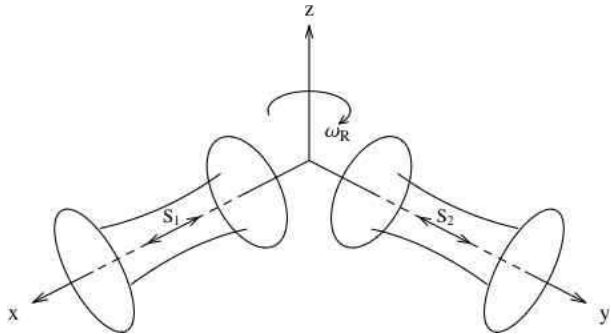}
\caption{Typical configuration of a rotating Lorentz invariance test using Fabry-Perot cavities.}
\label{FP}
\end{center}
\end{figure}

\begin{figure}[htbp]
\begin{center}
\includegraphics[width=10cm]{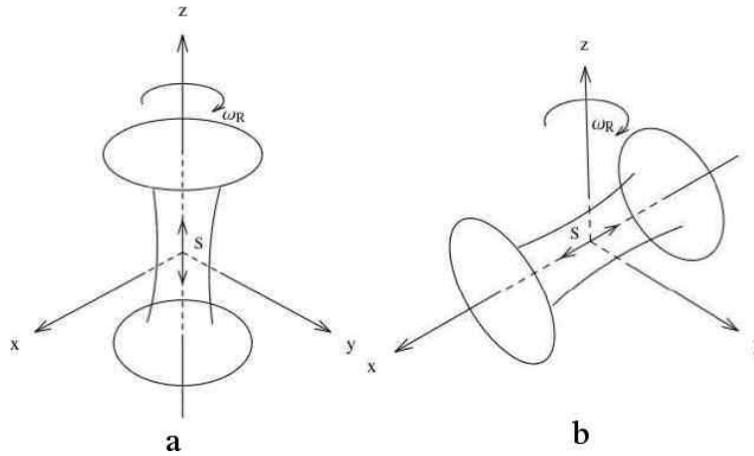}
\caption{New proposed matter filled Fabry-Perot cavity configurations in which two modes of orthogonal polarization are excited and compared, and are sensitive to violations in Lorentz invariance in the photon sector of the SME}
\label{FP2}
\end{center}
\end{figure}

Experiments based on laser stabilized Fabry-Perot resonators typically use either one \cite{Brillet} or two \cite{Muller} cavities placed with the lengths orthogonal to the laboratory $z$-axis. In a vacuum filled cavity it is easy to show that $|\mathcal{S}| = \frac{1}{2}$ for the configuration in Fig .\ref{FP}. In contrast, when one rotating cavity is compared to a stationary one the value is reduced by a factor of 2, to $|\mathcal{S}| = \frac{1}{4}$. 
It is also interesting to consider the sensitivity of matter filled cavities in the photon sector. Here, for simplicity we assume the relative permeability and permittivity are scalars of $\mu_r$ and $\epsilon_r$ respectively. It is straight forward to add anisotropy and only modifies the sensitivity slightly, so for brevity is not considered here. If similar configurations to Fig. \ref{FP} are constructed from solid material the sensitivity factor, $\mathcal{S}$, becomes dependent on polarization. This effect also allows for a sensitive experiment by exciting two modes of different polarization inside one cavity (Dual-Mode), of which some examples are shown in Fig . \ref{FP2}. Such cavities have been built previously to measure birefringence \cite{Hall}. High finesse matter cavities can be made by using low-loss crystalline dielectric materials at optical frequencies \cite{FSC,Schiller}. The sensitivity for different configurations are compared in Table \ref{fpcomp}. 
\begin{table}[htdp]
\begin{center}
\begin{tabular}{|c|c|}
\hline
Configuration & Sensitivity Factor $\mathcal{S}$ \\
\hline
\hline
Fig . \ref{FP} $E_z$ & $\frac{\mu_r}{2}$ \\
Fig . \ref{FP} $H_z$ & $\frac{1}{2\epsilon_r}$\\
Fig . \ref{FP} Circular Polarization & $\frac{1}{4}\left(\frac{1}{\epsilon_r}+\mu_r\right)$ \\
Fig . \ref{FP2} (a) & $\frac{1}{2}\left(\frac{1}{\epsilon_r}-\mu_r\right)$\\
Fig . \ref{FP2} (b) & $\frac{1}{4}\left(\frac{1}{\epsilon_r}-\mu_r\right)$\\
\hline
\end{tabular}
\end{center}
\caption{Value of the $\mathcal{S}$ factor for various configurations of Fabry-Perot cavity experiments}
\label{fpcomp}
\end{table}
For a low-loss dielectric material with $E_z$ polarization in the two orthogonal cavities (Fig .\ref{FP}) the sensitive factor, $\mathcal{S}$, is the same as the vacuum cavity, while for the circularly polarized case, the sensitivity is close to that of the single vacuum cavity resonator. In contrast the same experiment with $H_z$ polarization has reduced sensitivity of the order of the permittivity of the material. The sensitivity of the two Dual-Mode resonators gives the possibility of realizing a similar sensitivity to dual cavity experiments, but within the same cavity. The configuration should have a large degree of common mode rejection, and will be much more insensitive to external effects like temperature, vibration etc. and other systematics, and may be worth pursuing for these reasons. Note that M\"uller has recently completed an analysis of conventional cavity configurations in the electron (due to dispersion changes) and photon sector \cite{MulMat}. In our analysis we have only considered the photon sector and we have proposed some new unconventional configurations. It may be interesting to analyze these configurations in the electron sector. In the next subsection we consider similar configurations for Whispering Gallery (WG) modes.

\subsection{Whispering gallery mode resonators}
\label{WGr}
\begin{figure}[htdp]
\begin{center}
\includegraphics[width=8cm]{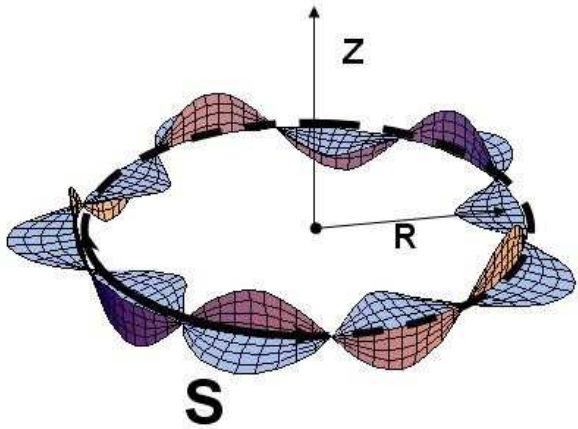}
\caption{A visual representation of the electric and magnetic fields 
of a 'pure' Whispering Gallery mode propagating in the $\phi$ direction with a radius of $r=R$.}
\label{wg-diagram}
\end{center}
\end{figure}

In this subsection we consider 'pure' WG modes, 
with the electric and magnetic fields propagating around with cylindrical symmetry at a discontinuity, with the direction of the Poynting vector ($\mathbf{E} \times \mathbf{B}$) as shown in figure \ref{wg-diagram}. Thus, it is natural to analyse such modes in cylindrical coordinates $\{r,\phi,z\}$.

For an actual WG mode the wave is reflected off an electromagnetic discontinuity, and the fields mainly lie within the radius of the discontinuity and a smaller inner caustic\cite{WolfGRG}. However, by taking the limit as the azimuthal mode number $m$ tends to infinity, the inner caustic converges to the  radius of the discontinuity and the fields are reduced to a Dirac delta function. There are two possible polarizations, WGE with dominant $H_z$ and $E_r$ fields and WGH with dominant $E_z$ and $H_r$. For 'pure' WG modes, WGE have non-zero electric and magnetic filling factors of $Pe_r = 1$ and $Pm_z = 1$, and WGH have electric and magnetic filling factors of $Pe_z = 1$ and $Pm_r = 1$, in cylindrical coordinates. The electric and magnetic filling factors may be converted from cylindrical to cartesian symmetry by (the $z$ component of the filling factor need not be transformed):
\begin{equation}
\label{filling factor average}
Pe_x = Pe_y = \frac{Pe_r+Pe_\phi}{2} \ \ : \ \ Pm_x = Pm_y = \frac{Pm_r+Pm_\phi}{2}
\end{equation}

\begin{figure}[htbp]
\begin{center}
\includegraphics[width=8cm]{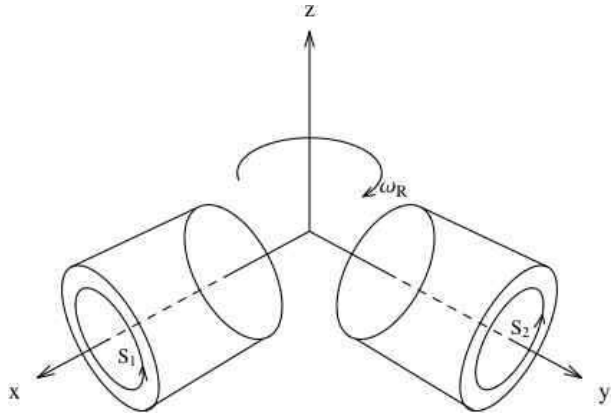}
\caption{Rotating Lorentz invariance test using two WG mode cavities}
\label{WG}
\end{center}
\end{figure}

\begin{figure}[htbp]
\begin{center}
\includegraphics[width=10cm]{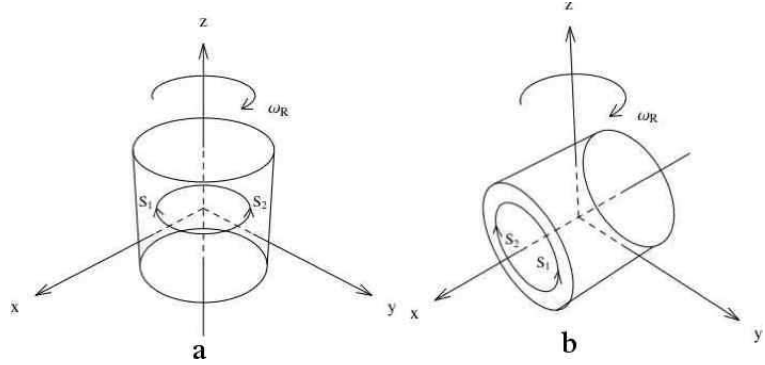}
\caption{Rotating Lorentz invariance test using two WG modes excited in one cavity (Dual-Mode resonator)}
\label{WG2}
\end{center}
\end{figure}

We can now do a similar analysis to subsection \ref{FPr} for configurations shown in  Fig. \ref{WG} and \ref{WG2} for the WG case, with the computed sensitivities listed in Table \ref{wgcomp}. In vacuum the $\mathcal{S}$ factor is half that of the FP cavities in subsection \ref{FPr}, and the Dual-Mode resonator is insensitive. However, in a low loss dielectric the $\mathcal{S}$ factor approaches the same value for WGE modes as the FP cavity experiments, but the WGH modes remain about a factor of two less sensitive. The value of the $\mathcal{S}$ factor for the Dual-Mode resonator is the mean value of the WGE and WGH modes.

\begin{table}[htdp]
\begin{center}
\begin{tabular}{|c|c|}
\hline
Configuration & Sensitivity Factor $\mathcal{S}$ \\
\hline
\hline
Fig . \ref{WG} $WGH$ &$-\frac{1}{2\epsilon_r}+\frac{\mu_r}{4}$\\
Fig . \ref{WG} $WGE$ &$\frac{1}{4\epsilon_r}-\frac{\mu_r}{2}$\\
Fig . \ref{WG2} (a) &$0$\\
Fig . \ref{WG2} (b) &$\frac{3}{8}\left(\frac{1}{\epsilon_r}-\mu_r\right)$\\
\hline
\end{tabular}
\end{center}
\caption{Value of the $\mathcal{S}$ factor for various configurations of WG mode resonator cavity experiments}
\label{wgcomp}
\end{table}

We have shown that similar sensitivities can be achieved with FP and WG cavity resonators. At UWA we have developed an experiment that uses low loss sapphire crystals, which exhibit a small uniaxial dielectric anistropy. The calculations of the sensitivity are presented section \ref{UWA}.

\section{Applying the RMS to Whispering Gallery Mode Resonator Experiments}
\label{RMSExp}

In this section we restrict ourselves to analysis of whispering gallery mode resonator experiments, as the analysis has been well described for Fabry-Perot resonators previously \cite{Muller2}. For the whispering gallery mode experiment as shown in Fig. \ref{WG}, the variable of integration around the path of the resonator is naturally chosen as the azimuthal angle, $\phi_j$, relative to the cylindrical co-ordinates of each resonator. Thus, from Eqn. (\ref{RMSunit}) a frequency shift due to a putative Lorentz
violation in the RMS framework is given by,
\begin{equation}
\frac{\Delta\nu_0}{\nu_0} = \frac{P_{MM}}{2\pi
c^2}\left[\oint{\left(\bf{v}.\hat{\bf{I}}_{a}(\phi_a)\right)^2
d\phi_a}-\oint{\left(\bf{v}.\hat{\bf{I}}_{b}(\phi_b)\right)^2
d\phi_b}\right] \label{RMS}
\end{equation}
The dominant components of the velocity vector $\bf{v}$ were already calculated in section \ref{RMSFW}, so to complete the calculation the time dependence of $\hat{\bf{I}}_{a}$ and $\hat{\bf{I}}_{b}$ must be calculated with respect to the MM-Earth frame. This of course depends on the sidereal and semi-sidereal frequencies, as well as the rotation frequency of the experiment. To start the calculation we define the time, $t=0$ when the axis of the two WG resonators are aligned as shown in Fig. \ref{WG} (i.e. the resonators align with the laboratory frame). Then from this time we assume the resonator is rotated in a anti-clockwise direction of frequency $\omega_s$, so the angle of rotation is $\gamma=\omega_s(t-t_s)$. Also, the longitudinal angle of the experiment is $\lambda$, which is dependent on the sidereal frequency and given by $\lambda=\omega_\oplus(t-t_l)$. Then we define the resonator with its cylinder axis in the $y$ direction as resonator $a$, and the resonator with its cylinder axis in the $x$ direction as resonator $b$. We also assume the WG modes are oscillating in a clockwise direction. In actual fact the calculation has been verified to be independent of the WG mode direction, and in most experiments is usually a standing wave (depending on the excitation) \cite{WGTW}. Thus in the laboratory frame at $t=0$ the unit vectors in the direction of the Poynting vector are;
\bea
	 {\bf{I}}_{a}(\phi_a)=\left(\begin{array}{ccc}
			-{\rm sin}\phi _a \\
			0  \\
			{\rm cos}\phi _a
		\end{array}
	 \right) \ \
	 {\bf{I}}_{b}(\phi_b)=\left(\begin{array}{ccc}
			0 \\
			{\rm cos}\phi _b  \\
			{\rm sin}\phi _b
		\end{array}
	 \right)
	 \label{I}
\eea
Now if we transform from the resonator to the laboratory, then to the MM-Earth frame the unit vectors become.
\bea
{\bf{I}}_{Earth:a}=\left(\begin{array}{ccc}
			{-\rm sin}\phi _a({\rm cos}\lambda{\rm cos}\chi{\rm cos}\gamma-{\rm sin}\gamma{\rm sin}\lambda)+{\rm cos}\phi _a{\rm cos}\lambda{\rm sin}\chi \\
			-{\rm sin}\phi _a({\rm cos}\lambda{\rm sin}\gamma+{\rm cos}\gamma{\rm cos}\chi{\rm sin}\lambda)+{\rm cos}\phi _a{\rm sin}\lambda{\rm sin}\chi   \\
			{\rm cos}\chi{\rm cos}\phi _a+{\rm sin}\chi {\rm cos}\gamma{\rm sin}\phi _a
		\end{array}
	 \right)
	 	 \label{I1} \\
{\bf{I}}_{Earth:b}=\left(\begin{array}{ccc}
			-{\rm sin}\phi _b({\rm cos}\lambda{\rm cos}\chi{\rm sin}\gamma+{\rm cos}\gamma{\rm sin}\lambda)+{\rm cos}\phi _b{\rm cos}\lambda{\rm sin}\chi \\
			{\rm sin}\phi _b({\rm cos}\lambda{\rm cos}\gamma-{\rm sin}\gamma{\rm cos}\chi{\rm sin}\lambda)+{\rm cos}\phi _b{\rm sin}\lambda{\rm sin}\chi   \\
			{\rm cos}\chi{\rm cos}\phi _b+{\rm sin}\chi {\rm sin}\gamma{\rm sin}\phi _b
		\end{array}
	 \right)  	 
	 \label{I2}
\eea
Here as in the previous sections $\chi$ is the angle from the north pole (co-latitude).

The next step is to substitute (\ref{I1}), (\ref{I2}) and (\ref{numvelocity}) into (\ref{RMS}). However, to be consistent with the SME analysis the phase should be calculated with respect to the vernal equinox, so that $\lambda_0=\Phi_0+\pi$ is substituted into Eqn. (\ref{numvelocity}) before we substitute it into (\ref{RMS}) to calculate the frequency shift. Also, because we defined the rotation to be clockwise in the SME, to be consistent we define $\gamma_R=\omega_R(t-t_s)$ where $\omega_R=-\omega_s$. In this case the frequency components, which experience a frequency shift are given in Table \ref{tabMM}.
\begin{table}
\begin{center}
\begin{tabular}{|c||c|}
\hline
$\omega_i$ & $10^7 Cu_{\omega_i}/P_{MM}$ \\
\hline
\hline
$2\omega_R$&$-3.904 + 3.904{\cos (\chi )}^2 + 0.098{\sin (\chi )}^2 $\\
& $-  \sin (\Phi_0) \left( -0.607 + 0.607{\cos (\chi )}^2 \right)$\\
& $- \cos (\Phi_0)\left( -0.120 + 0.120{\cos (\chi )}^2 - 0.055{\sin (\chi )}^2 \right)$\\
\hline
$2\omega_R +\omega_\oplus$&$-0.876\sin (\chi ) + 0.876\cos (\chi )\sin (\chi )$\\
& $+ \sin (\Phi_0)\left( 0.068\sin (\chi ) - 0.068\cos (\chi )\sin (\chi ) \right)$\\
& $+  \cos (\Phi_0)\left( -0.232\sin (\chi ) + 0.232\cos (\chi )\sin (\chi ) \right)$\\
\hline
$2\omega_R -\omega_\oplus$&$0.876\sin(\chi) + 0.876\cos (\chi )\sin(\chi )$\\
& $+\sin(\Phi_0)\left( -0.068\sin(\chi) - 0.068\cos(\chi)\sin(\chi)\right)$\\
& $+ \cos (\Phi_0)\left( 0.232\sin(\chi) + 0.232\cos(\chi)\sin(\chi)\right)$\\
\hline
$2\omega_R +2\omega_\oplus$&$1.952- 3.904\cos (\chi ) + 1.952{\cos (\chi )}^2 $\\
& $+\cos (\Phi_0)\left( -0.060 + 0.120\cos (\chi ) -0.060{\cos (\chi )}^2 \right)$\\
& $+ \sin (\Phi_0)\left( -0.303+ 0.607\cos (\chi ) -0.303{\cos (\chi )}^2 \right)$\\
\hline
$2\omega_R -2\omega_\oplus$&$1.952 + 3.904\cos (\chi ) + 1.952{\cos (\chi )}^2$\\
& $+ \cos (\Phi_0)\left( -0.060 - 0.120\cos (\chi ) - 0.060{\cos (\chi )}^2 \right)$\\
& $+ \sin (\Phi_0) \left( -0.303 - 0.607\cos (\chi ) - 0.303{\cos (\chi )}^2 \right)$\\
\hline
\end{tabular}
\end{center}
\caption{Dominant coefficients in the RMS, using a short data set approximation calculated from Eqn. (\ref{RMS}). }
\label{tabMM} 
\end{table}
From the results of the calculation we note that perturbations due to Lorentz violations occur at the same frequencies as the SME (see subsection \ref{RMSFWss}). Fortunately, it is not necessary to consider  perturbations at exactly twice the spin frequency, $2\omega_s$, that are primarily due to the larger systematic effects associated with the rotation, as we only need to put a limit on one parameter. Also, the cosine components $(Cu_{\omega_i})$ with respect to the CMB are the most sensitive, so we need not consider the sine components.

\subsection{Phase with respect to the CMB}

To extract the $P_{MM}$ term
from our data we must first determine the phase of our experiment
with respect to the CMB.  Thus, in similar way to the reasoning for the SME (see subsection \ref{phSCECF})  we require $\delta_R$, the difference
between the experiment's rotation time and the time since the vernal
equinox, and $\delta_{\oplus}$, the difference between the sidereal
time and the time since the vernal equinox.

\begin{figure}[h]
\begin{center}
\includegraphics[width=7cm]{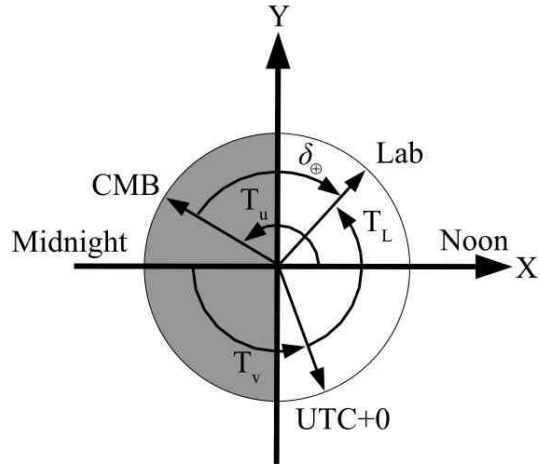}
\caption{Diagram showing meridians and angles used to determine the
phase of the experiment with respect to the CMB.} 
\label{RMSphase}
\end{center}
\end{figure}

As was the case for the SME, $\delta_R=0$ since the axis of symmetry of the first resonator, $a$, is aligned with the local $y$-axis at $T_R=0$. However, $\delta_{\oplus}$ will be different since in the case of the RMS it is measured with respect to the CMB (or MM-Earth frame), not the SCCEF. The CMB is oriented at 11.2 h right ascension, 6.4 degrees declination
relative to the equatorial plane. Let us define $T_v$ to be the time
in seconds after midnight UTC+0, at which the vernal equinox has
occurred in the J2000.0 frame \cite{KM}. $T_u$ is the direction of
the CMB (11.2h). For convenience we also define our longitude $T_L$
in terms of sidereal seconds from midnight (in the case of our
laboratory $T_l = 115.826^\circ \times \frac{23 hr 56
min}{360^\circ} = 27721 sec$). As shown in Fig. \ref{RMSphase}, we
now have an expression for $\delta_{\oplus}$.
\begin{equation}
\delta_{\oplus} = T_L + T_v - (T_u + \frac{23hr56min}{2})
\end{equation}

Hence we are able to determine the phase of the experiment orientation relative to the CMB.

\section{The University of Western Australia Rotating Experiment}
\label{UWA}

\begin{figure}
\begin{center}
\includegraphics[width=3in]{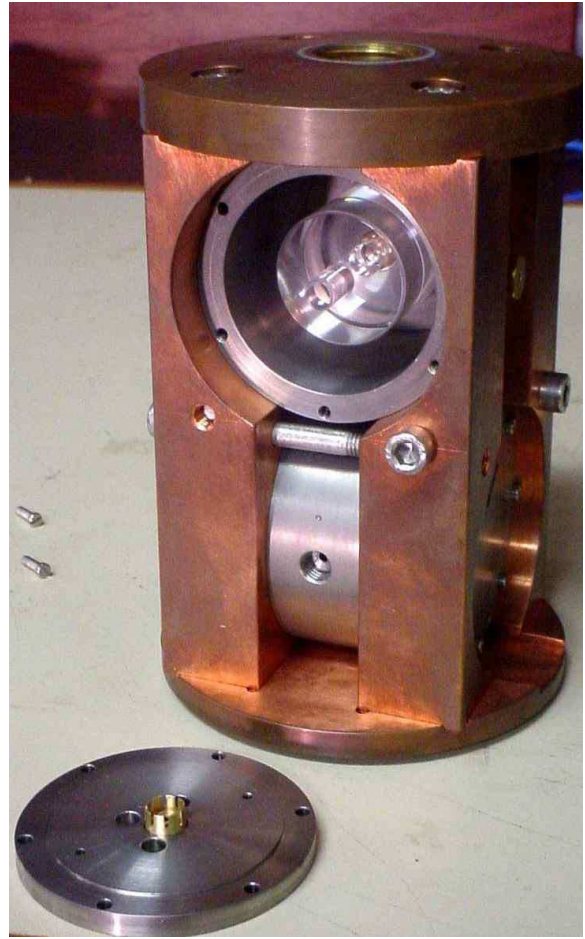}
\caption{The two resonators are positioned orthogonal to each other in the
mounting structure. One of the sapphires can be seen mounted inside the
superconducting niobium cavity. The spindles are firmly held in each lid by
sprung brass bushes.} \label{fig:slc}
\end{center}
\end{figure}

\begin{figure}
\begin{center}
\includegraphics[width=3.5in]{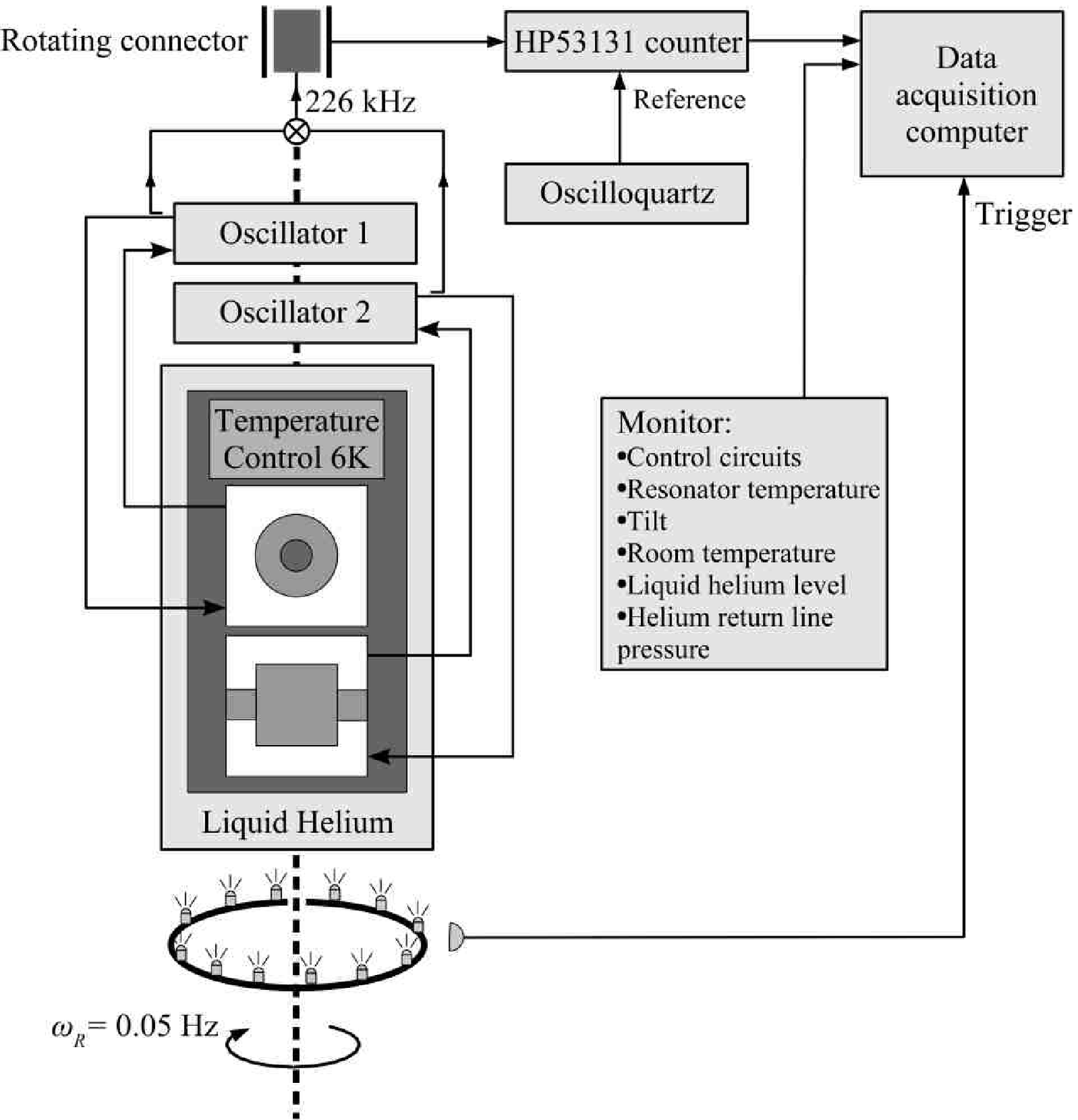}
\caption{Schematic of the experimental setup} \label{fig:setup}
\end{center}
\end{figure}

Our experiment consists of two cylindrical sapphire resonators of 3 cm diameter and height supported by spindles at either end within superconducting niobium cavities \cite{Giles}, which are oriented with their
cylindrical axes orthogonal to each other in the horizontal plane (see Fig. \ref{fig:slc}).
Whispering gallery modes \cite{wgmode} are excited close to 10 GHz, with a difference frequency of 226 kHz. The frequencies are
stabilized using a Pound locking scheme, and amplitude variations
are suppressed using an additional control circuit. A detailed
description of the cryogenic oscillators can be found in
\cite{Mann, Hartnett}, and a schematic of the experimental setup is shown in Fig. \ref{fig:setup}. The resonators are
mounted in a common copper block, which provides
common mode rejection of temperature fluctuations due to high
thermal conductivity at cryogenic temperatures. The structure is in
turn mounted inside two successive stainless steel vacuum cylinders
from a copper post, which provides the thermal connection between the
cavities and the liquid helium bath. A stainless steel section
within the copper post provides thermal filtering of bath
temperature fluctuations. A foil heater and carbon-glass temperature
sensor attached to the copper post controls the temperature set point
to 6 K with mK stability. Two stages of vacuum isolation are
used to avoid contamination of the sapphire resonators from the
microwave and temperature control devices located in the cryogenic
environment.
\begin{figure}
\begin{center}
\includegraphics[width=3.5in]{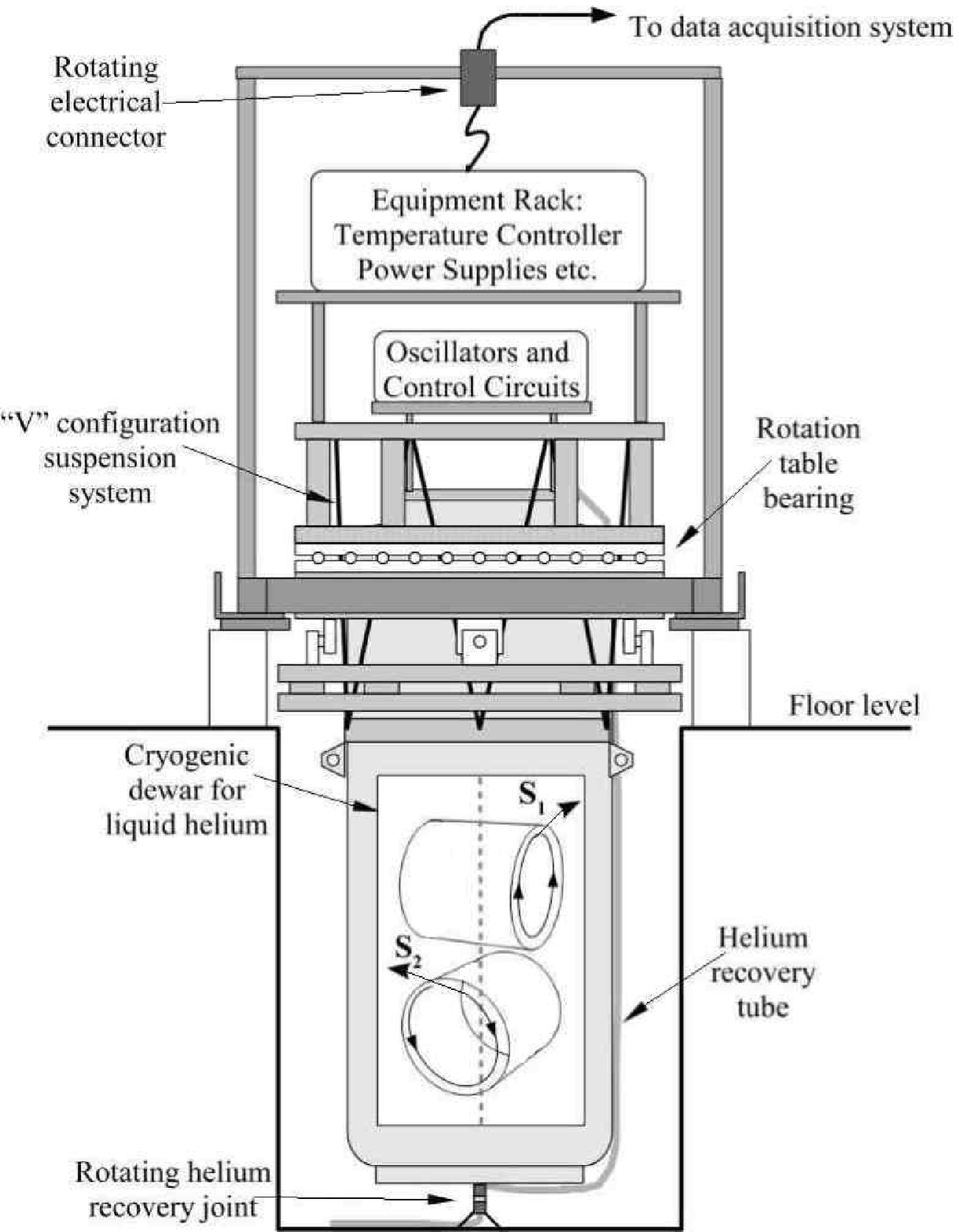}
\caption{Schematic of the cryogenic dewar, mounted in the rotation
table. Inside the dewar a schematic of the two orthogonally
orientated resonators is shown, along with the Poynting vectors of
propagation $S_{1}$ and $S_{2}$.} \label{fig:rotPic}
\end{center}
\end{figure}

A schematic of the rotation system is shown in Fig.\ref{fig:rotPic}.
A cryogenic dewar containing the resonators, along with the room
temperature oscillator circuits and control electronics, is
suspended within a ring bearing. A multiple "V" shaped suspension
made from loops of elastic shock cord avoids high Q-factor pendulum
modes by ensuring that the cord has to stretch and shrink (providing
damping losses) for horizontal motion as well as vertical. The
rotation system is driven by a microprocessor controlled stepper
motor. A commercial 18 conductor slip ring connector, with a hollow through
bore, transfers power and various signals to and from the rotating
experiment. A mercury based rotating coaxial connector
transmits the difference frequency to a stationary frequency counter
referenced to an Oscilloquartz oscillator. The data acquisition
system logs the difference frequency as a function of orientation,
as well as monitoring systematic effects including the temperature
of the resonators, liquid helium bath level, ambient room
temperature, oscillator control signals, tilt, and helium return
line pressure.

Inside the sapphire crystals standing waves are set up with the dominant electric and
magnetic fields in the axial and radial directions respectively,
corresponding to a propagation (Poynting) vector around the
circumference. The observable of the experiment is the difference
frequency, and to test for Lorentz violations the perturbation of
the observable with respect to an alternative test theory
must be derived. The mode which we excite is a Whispering Gallery mode 
we have a choice of WGE$_{m,n,p}$ or WGH$_{m,n,p}$ modes, the first subscript, $m$, gives the azimuthal mode number, while $n$ and $p$ give the 
number of zero crossings in the radial and z-direction respectively. Typically the so called fundamental mode families WGE$_{m,0,0}$ or WGH$_{m,0,0}$ as they have the highest Q-factors.  To calculate the sensitivity in the RMS we use the technique presented in section \ref{RMSExp}, while in the following subsection we numerically compute the sensitivity in the SME.

\subsection{Sensitivity in the SME}

\begin{figure}
\begin{center}
\includegraphics[width=8cm]{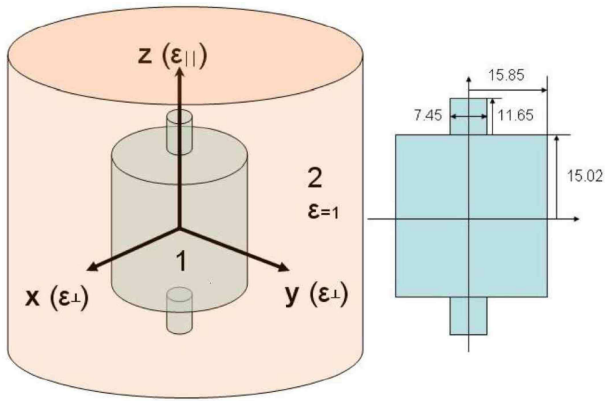}
\caption{Schematic of a cylindrical sapphire crystal resonator, with dimensions in mm. The crystal exhibits uniaxial anisotropy with the axis of symmetry (c-axis) aligned along the cylindrical $z$ axis.  The permittivity along the c-axis given by $\epsilon_{||}$. Perpendicular to the c-axis in the $x$, $y$ or $r$, $\phi$ plane, the permittivity is given by $\epsilon_\bot$. The two regions shown are 1. the crystal and 2. the cavity.}
\label{crystal-epsilon}
\end{center}
\end{figure}

\begin{figure}[hbt]
\begin{center}
\includegraphics[width=10cm]{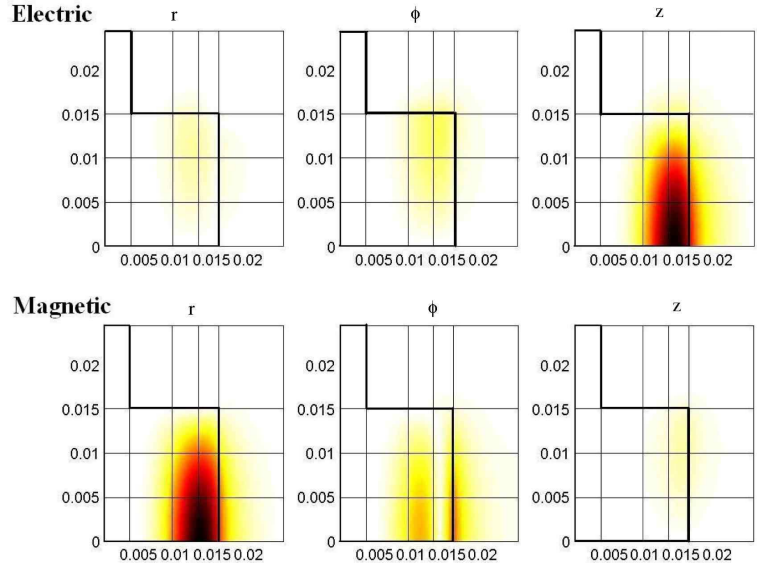}
\caption{The magnetic and electric field density plots shown in the 
top right-hand quadrant of an axial slice through the sapphire 
crystal and cavity for the $WGH_{8,0,0}$ mode. Note the dominant fields 
are $E_z$ and $H_r$ consistent with a pure whispering gallery mode approximation.}
\label{field-E811}
\end{center}
\end{figure}

In this subsection we calculate the sensitivity of the fundamental WG mode families, WGE$_{m,0,0}$ and WGH$_{m,0,0}$ to putative Lorentz violation in the SME, and compare it with the 'pure' WG approximation given in Fig. \ref{WG}. For a proper analysis of the sapphire loaded cavity resonators two regions of space need to be taken into account: the anisotropic crystal and the cavity free space surrounding it (see Fig. \ref{crystal-epsilon}). The latter has a relative permittivity of 1, while both have relative permeability of 1 in all directions. The calculations proceed by splitting up $V$ into $V_1$ (the crystal) and $V_2$ (the freespace), so we may sum the components of the $\mathcal{M}$ matricies over the two volumes (see Eqns. (\ref{scriptmDE}) and (\ref{scriptmHB})). The resonator operates close to liquid helium temperatures (6 Kelvin), where the permittivity of sapphire is, $\epsilon_\bot = 9.272$ and $\epsilon_{||} = 11.349$.
To determine the sensitivity, we need to just calculate the experiments $\mathcal{S}$ factor in a similar way to the calculation for the 'pure' WG modes in subsection \ref{WGr}. In this case the electric and magnetic filling factors must be calculated using a numeric technique such as finite element analysis, method of lines or separation of variables\cite{WolfGRG}. In this work we have chosen to use method of lines developed at IRCOM at the University of Limoges  \cite{MoL} . The calculated electric and magnetic field densities for the chosen mode $(WGH_{8,0,0})$ of operation at 10 GHz is shown in figure \ref{field-E811}, and the $\mathcal{S}$ factor is calculated to be 0.19575.
\begin{figure}[ht]
\begin{center}
\includegraphics[width=12cm]{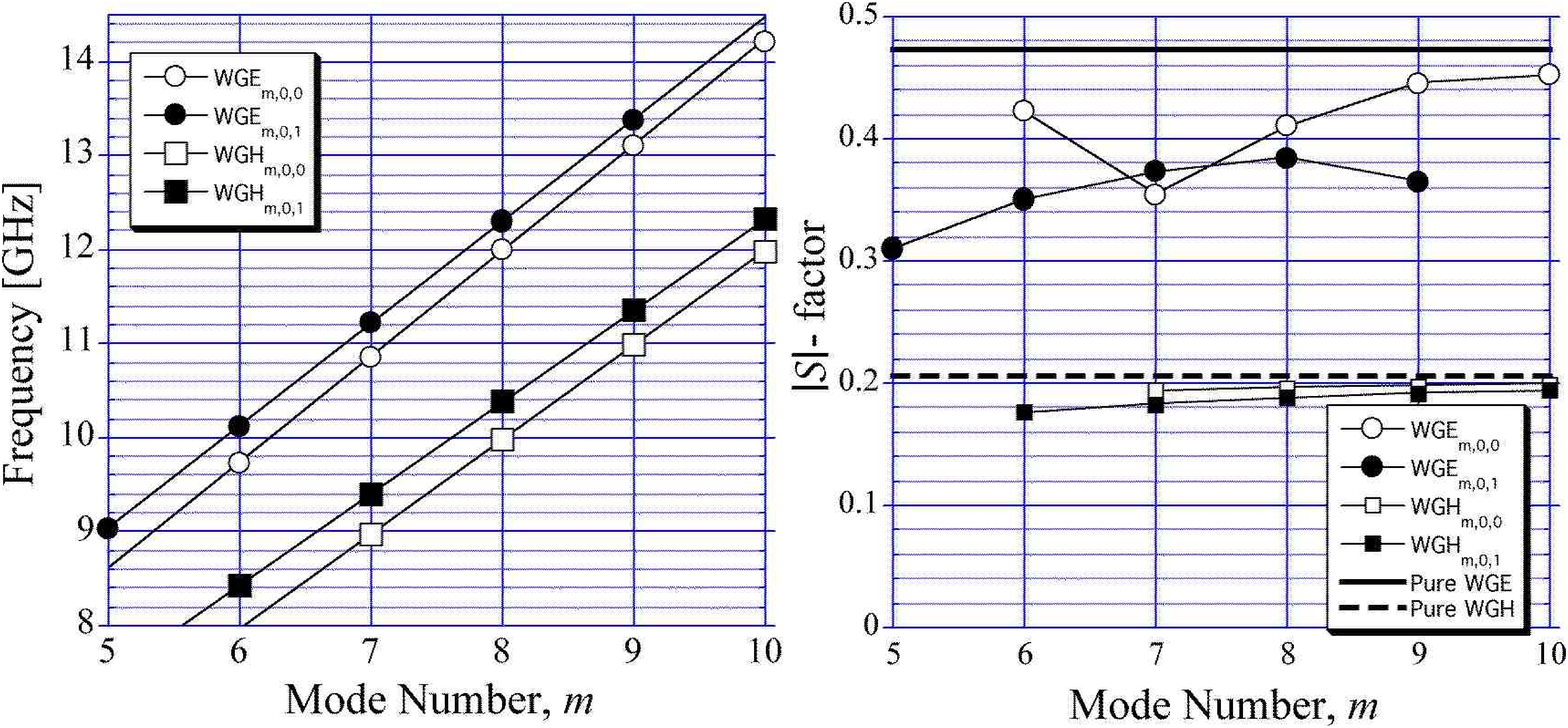}
\caption{The frequency and $|\mathcal{S}|$ factor as a function of mode number, $m$ for the two lowest frequency $WGE$ and $WGH$ mode families. Note the general convergence to the 'pure' WG value as $m$ increases.}
\label{total-sense}
\end{center}
\end{figure}

The actual WG modes have all field components in both regions of the crystal. This 
modifies the sensitivity slightly, but approaches the limit of the 'pure' WG mode as $m \rightarrow 
\infty$. The magnitude of the $\mathcal{S}$ factor for the fundamental $WGE$ and $WGH$ modes at  X-Band (8GHz-12GHz) are plotted in Fig. \ref{total-sense}. The WGH modes seem converge nicely towards the predicted 
'pure' WGH mode sensitivity, while the WGE modes have a dip in 
sensitivity. This can be explained by an intersection 
with another mode of the same $m$ number, 
resulting in a spurious mode interaction\cite{whispering} . This does not
occur in WGH modes since they are the lowest frequency modes for the mode number $m$ (refer to figure 2 of \cite{whispering}). It is important to note that about a factor of two in sensitivity can be gained if we use a WGE mode rather than a WGH mode. However because we are using a 3 cm crystal rather than a 5 cm crystal, the Q-factor of WGE modes are degraded due to radiation and wall losses. In the future we can markedly improve the sensitivity by employing the typical 5 cm cavities that operate in WGE modes, as were used in the non-rotating experiments \cite{WolfGRG}.

\section{Data Analysis and Interpretation of Results}
\label{DA}

\begin{figure}
\begin{center}
\includegraphics[width=3.5in]{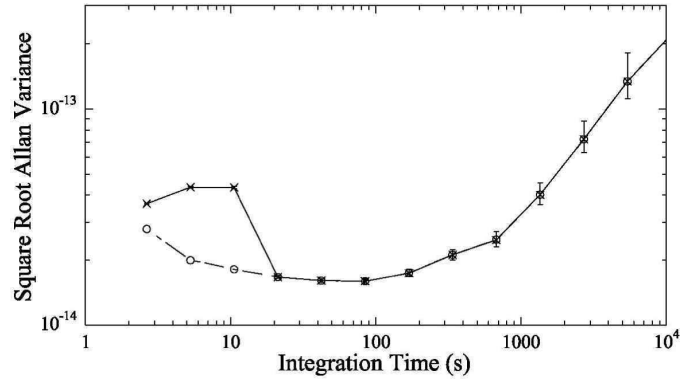}
\caption{Square Root Allan Variance fractional frequency instability measurement of the difference frequency when rotating (crosses) and stationary (circles). The hump at short integration times is due to systematic effects associated with the rotation of the experiment, with a period of 18 seconds. Above 18 seconds the instability is the same as when the experiment is stationary.}
\label{fig:stab}
\end{center}
\end{figure}

 Fig.\ref{fig:stab} shows typical fractional frequency instability of the 226 kHz difference with respect to 10 GHz, and compares the instability when rotating and stationary. A minimum of $1.6\times10^{-14}$ is recorded at 40s. Rotation induced systematic effects degrade the stability up to 18s due to signals at the rotation frequency of $0.056 Hz$ and its harmonics. We have determined that tilt variations dominate the systematic by measuring the magnitude of the fractional frequency dependence on tilt and the variation in tilt at twice the rotation frequency, $2\omega_R (0.11 Hz)$, as the experiment rotates. We minimize the effect of tilt by manually setting the rotation bearing until our tilt sensor reads a minimum at $2\omega_R$. The latter data sets were up to an order of magnitude reduced in amplitude as we became more experienced at this process. The remaining systematic signal is due to the residual tilt variations, which could be further annulled with an automatic tilt control system. It is still possible to be sensitive to Lorentz violations in the presence of these
systematics by measuring the sidereal, $\omega_\oplus$, and semi-sidereal, $2\omega_\oplus$, sidebands about $2\omega_R$, as was done in \cite{Brillet}. The
amplitude and phase of a Lorentz violating signal is determined by fitting the parameters of Eq. (\ref{nuTest}) to the data, with the phase of the fit adjusted according to the test theory used.
\begin{equation}
\frac{\Delta\nu_0}{\nu_0} = A + B t + \sum_i C_{\omega_i} {\rm
cos}(\omega_{i}t + \varphi_i) + S_{\omega_i} {\rm sin}(\omega_{i}t +
\varphi_i)
\label{nuTest}
\end{equation}
Here $\nu_0$ is the average unperturbed frequency of the two
sapphire resonators, and  $\Delta\nu_0$ is the perturbation of the
226 kHz difference frequency, $A$ and $B$ determine the frequency offset
and drift, and $C_{\omega_i}$ and $S_{\omega_i}$ are the amplitudes of a cosine and
sine at frequency $\omega_i$  respectively. In the final analysis we
fit 5 frequencies to the data, $\omega_i = (2\omega_R,
2\omega_R\pm\omega_\oplus, 2\omega_R\pm 2\omega_\oplus)$, as well as
the frequency offset and drift. The correlation coefficients between
the fitted parameters are all between $10^{-2}$ to $10^{-5}$. 
Since the residuals exhibit a significantly non-white behavior, 
the optimal regression method is weighted least squares (WLS) \cite{Wolf04}. 
WLS involves pre-multiplying both the experimental data and the model matrix by a
whitening matrix determined by the noise type of the residuals of an ordinary least squares analysis.

\begin{figure}
\begin{center}
\includegraphics[width=8cm]{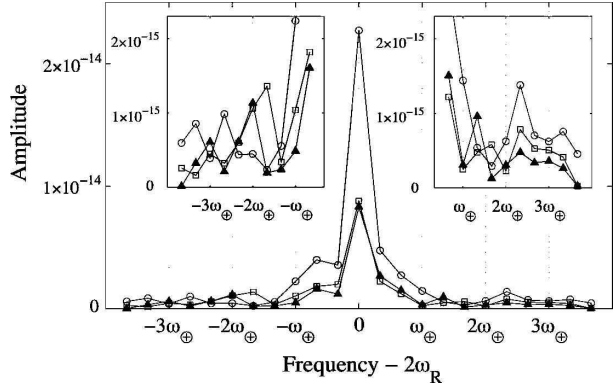}
\caption{Spectrum of amplitudes $\sqrt{C_{\omega_i}^2+S_{\omega_i}^2}$ calculated using WLS, showing systematic leakage about $2\omega_R$ for 2 data sets, data set 1 (3.6 days, circles), data set 5 (6.1 days, squares) and the combined data (18 days spanning 3 months, solid triangles). Here $\omega_\oplus$ is the sidereal frequency $(11.6 \mu Hz)$. By comparing a variety of data sets we have seen that leakage is reduced in longer data sets with lower systematics. The insets show the typical amplitude away from the systematic, which have statistical uncetainties of order $10^{-16}$.}
\label{fig:data}
\end{center}
\end{figure}

We have acquired 5 sets of data over a period of 3 months beginning
December 2004, totaling 18 days.  The length of the sets (in days)
and size of the systematic are ($3.6, 2.3\times10^{-14}$), ($2.4,
2.1\times10^{-14}$), ($1.9, 2.6\times10^{-14}$), ($4.7, 1.4 \times
10^{-15}$), and ($6.1, 8.8 \times 10^{-15}$) respectively. We have
observed leakage of the systematic into the neighboring side bands
due to aliasing when the data set is not long enough or the
systematic is too large. Fig.\ref{fig:data} shows the total
amplitude resulting from a WLS fit to 2 of the data sets over a
range of frequencies about $2\omega_R$. It is evident that the
systematic of data set 1 at $2\omega_R$ is affecting the fitted
amplitude of the sidereal sidebands $2\omega_R\pm\omega_\oplus$ due
to its relatively short length and large systematics. By analyzing
all five data sets simultaneously using WLS the effective length of
the data is increased, reducing the width of the systematic
sufficiently as to not contribute significantly to the sidereal and semi-sidereal
sidebands.

\subsection{Standard Model Extension Framework}
\label{SMEFWss}

In the photon sector of the SME 10 independent components of
$\tilde{\kappa}_{e+}$ and $\tilde{\kappa}_{o-}$ have been
constrained by astronomical measurements to $< 2\times 10^{-32}$
\cite{KM,Kost01}. Seven components of $\tilde{\kappa}_{e-}$ and
$\tilde{\kappa}_{o+}$ have been constrained in optical and microwave
cavity experiments \cite{Muller,Wolf04} at the $10^{-15}$ and
$10^{-11}$ level respectively, while the scalar
$\tilde{\kappa}_{tr}$ component recently had an upper limit set of
$< 10^{-4}$ \cite{TobarPRD}. The remaining
$\tilde{\kappa}_{e-}^{ZZ}$ component could not be previously
constrained in non-rotating experiments \cite{Muller,Wolf04}. In contrast, our rotating experiment is sensitive to $\tilde{\kappa}_{e-}^{ZZ}$. However, it appears only at $2\omega_R$, which is dominated by systematic effects. By using the formulas derived in Table \ref{approx} for short data sets and the $\mathcal{S}$ factor for the $WGH_{8,0,0}$ mode in Fig. \ref{total-sense}, the resulting numerical relation between the
parameters of the SME and the $C_{\omega_i}$ and $S_{\omega_i}$ coefficients were calculated and are given
in Table \ref{taba}.

From our combined analysis of all data sets, and using the relation to $\tilde{\kappa}_{e-}^{ZZ}$ given in Table \ref{taba}, we determine a value for $\tilde{\kappa}_{e-}^{ZZ}$
of $4.1(0.5)\times10^{-15}$. However, since we do not know if the
systematic has canceled a Lorentz violating signal at $2\omega_R$, we cannot reasonably claim this as an upper limit. Since we have five individual data sets, a limit can be set  by treating the $C_{2\omega_R}$ coefficient as a statistic. The phase of the systematic depends on the initial experimental conditions, and is random across the data sets. Thus, we have five values of $C_{2\omega_R}$, ($\{-4.2,11.4, 21.4, 1.3, -8.1\}$ in $10^{-15}$), two are negative coefficients and three are positive. If we take the mean of these coefficients, the systematic signal will cancel if the phase is random, but the possible Lorentz violating signal will not, since the phase is constant. Thus a limit can be set by taking the mean and standard deviation of the five coefficient of $C_{2\omega_R}$. This gives a more conservative bound of $2.1(5.7)\times 10^{-14}$, which includes zero. Our experiment is also sensitive to all other seven components of $\tilde{\kappa}_{e-}$ and $\tilde{\kappa}_{o+}$ (see Table \ref{taba}) and improves present limits by up to a factor of 7, as shown in Table \ref{tabb}.

\begin{table}[htdp]
\begin{center}
\begin{tabular}{|c||c|c|c|c|}
\hline
$\omega_i$ & $C_{\omega_i}$ & $S_{\omega_i}$ \\
\hline
\hline
$2\omega_R$ & $0.21\tilde{\kappa}_{e-}^{ZZ}$ & - \\
\hline
$2\omega_R + \omega_\oplus$ & $2.5\times 10^{-5}\sin\Phi_0\tilde{\kappa}_{o+}^{XY}$& $-\cos\Phi_0\left[2.3\times 10^{-5}\tilde{\kappa}_{o+}^{XY}-1.0\times 10^{-5}\tilde{\kappa}_{o+}^{XZ}\right]$ \\
& $-1.0\times 10^{-5}\cos\Phi_0\tilde{\kappa}_{o+}^{YZ}$ &$-0.27\tilde{\kappa}_{e-}^{YZ}$ \\
&$-0.27 \ \tilde{\kappa}_{e-}^{XZ}$& \\
\hline
$2\omega_R +2\omega_\oplus$ & $-2.1\times 10^{-5}\cos\Phi_0\tilde{\kappa}_{o+}^{XZ}$ \  \  \ & $-2.3\times 10^{-5}\sin\Phi_0\tilde{\kappa}_{o+}^{XZ}$ \\
&$+2.3\times 10^{-5}\sin\Phi_0\tilde{\kappa}_{o+}^{YZ}$ &$-2.1\times 10^{-5}\cos\Phi_0\tilde{\kappa}_{o+}^{YZ}-0.23 \ \tilde{\kappa}_{e-}^{XY}$\\
&$-0.11(\tilde{\kappa}_{e-}^{XX}-\tilde{\kappa}_{e-}^{YY})$& \\
\hline
$2\omega_R - \omega_\oplus$ & $-0.31C_{2\omega_R +\omega_\oplus}$ & $0.31S_{2\omega_R +2\omega_\oplus}$ \\
\hline
$2\omega_R -2\omega_\oplus$ & $9.4\times 10^{-2}C_{2\omega_R +2\omega_\oplus}$ & $-9.4\times10^{-2}S_{2\omega_R +2\omega_\oplus}$ \\
\hline
\end{tabular}
\end{center}
\caption{Coefficients $C_{\omega_i}$ and $S_{\omega_i}$ in (1) for the
five frequencies of interest and their relation to the components of
the SME parameters $\tilde{\kappa}_{e-}$ and $\tilde{\kappa}_{o+}$,
derived using a short data set approximation including terms up to
first order in orbital velocity, where $\Phi_0$ is the phase of the
orbit since the vernal equinox. Note that for short data sets the
upper and lower sidereal sidebands are redundant, which reduces the
number of independent measurements to 5. To lift the redundancy,
more than a year of data is required so annual offsets may be
de-correlated from the twice rotational and sidereal sidebands
listed.}
\label{taba} 
\end{table}
\begin{table}[htdp]
\begin{center}
\begin{tabular}{|c|c|c|c|c|}
\hline
  & $\tilde{\kappa}_{e-}^{XY}$ & $\tilde{\kappa}_{e-}^{XZ}$ & $\tilde{\kappa}_{e-}^{YZ}$ & $(\tilde{\kappa}_{e-}^{XX}-\tilde{\kappa}_{e-}^{YY})$ \\
\hline
this work & -0.63(0.43) & 0.19(0.37) & -0.45(0.37) & -1.3(0.9)\\
from \cite{Wolf04} & -5.7(2.3) & -3.2(1.3) & -0.5(1.3) & -3.2(4.6)\\
\hline
\hline 
 &  $\tilde{\kappa}_{e-}^{ZZ}$ & $\tilde{\kappa}_{o+}^{XY}$ & $\tilde{\kappa}_{o+}^{XZ}$ & $\tilde{\kappa}_{o+}^{YZ}$ \\
\hline
this work& $21(57)$ & 0.20(0.21) & -0.91(0.46) & 0.44(0.46) \\
from \cite{Wolf04}  & $-$ & -1.8(1.5) & -1.4(2.3) & 2.7(2.2) \\
\hline
\end{tabular}
\end{center}
\caption{Results for the SME Lorentz violation parameters, assuming no cancelation between the isotropy terms. $\tilde{\kappa}_{e-}$ (in $10^{-15}$) and first order boost terms $\tilde{\kappa}_{o+}$ (in $10^{-11}$) \cite{Lipa}.}
\label{tabb}
\end{table}

\subsection{Robertson, Mansouri, Sexl Framework}
\label{RMSFWss}
From Eqn. (\ref{RMS}), the dominant coefficients are calculated to be only due to the cosine terms with respect to the CMB right ascension, $Cu_{\omega_i}$, and the theory predicts no perturbations in the quadrature term. Since our experiment is rotating clock wise we can substitute $2\omega_s=-2\omega_R$, and once we perform the integral and substitute all the numeric values. Following this method we calculate the coefficients as shown in Table \ref{Tabrms}.
\begin{table}
\begin{center}
\begin{tabular}{|c||c|c|c|c|}
\hline
$\omega_i$ & $Cu_{\omega_i}$ & $P_{MM}$ \\
\hline
\hline
$2\omega_R +\omega_\oplus$&$[-1.13\times10^{-7}-3.01\times10^{-8}\cos\Phi_{0}$ & $-2.1(7.2)$\\
& $+8.83\times10^{-9}\sin\Phi_{0}]P_{MM}$ &\\
$2\omega_R -\omega_\oplus$&$[3.51\times10^{-8}+9.31\times10^{-9}\cos\Phi_{0}$ & $62.4(23.3)$\\
& $-2.73\times10^{-9}\sin\Phi_{0}]P_{MM}$ &\\
$2\omega_R +2\omega_\oplus$&$[4.56\times10^{-7}-1.39\times10^{-8}\cos\Phi_{0}$ & $-1.3(2.1)$\\
& $-7.08\times10^{-8}\sin\Phi_{0}]P_{MM}$ &\\
$2\omega_R -2\omega_\oplus$&$[4.37\times10^{-8}-1.34\times10^{-9}\cos\Phi_{0}$ & $-7.5(22.1)$\\
& $-6.78\times10^{-9}\sin\Phi_{0}]P_{MM}$ & \\
\hline
\end{tabular}
\end{center}
\caption{Dominant coefficients in the RMS, using a
short data set approximation calculated from Eq. (\ref{RMS}). The measured values of $P_{MM}$ (in $10^{-10}$) are shown together with
the statistical uncertainties in the bracket. From this data the
measured and statistical uncertainty of $P_{MM}$ is determined to be
$-0.9(2.0)\times10^{-10}$, which represents more than a factor of 7.5
improvement over previous results $2.2(1.5)\times10^{-9}$\cite{Muller}.}
\label{Tabrms} 
\end{table}

The same five data sets were then re-analysed in the correct quadrature with respect to the CMB, with the results listed with the coefficients in Table \ref{Tabrms}. The measured and statistical uncertainty of $P_{MM}$ is determined to be $-0.9(2.0)\times10^{-10}$, which represents a factor of 7.5 improvement over previous results $2.2(1.5)\times10^{-9}$\cite{Muller}.

\section{Summary}
\label{CD}

Rotating resonator experiments are emerging as one of the most sensitive types of Local Lorentz Invariance tests in electrodynamics (also see other contributions within these proceedings). In this work  we have analysed in detail such experiments to putative Lorentz violation in both the RMS and SME frameworks. In the RMS, rotating experiments only enhance the sensitivity to the Michelson-Morley parameter, $P_{MM}$, and are not sensitive to the Kennedy-Thorndike , $P_{KT}$, or Ives-Stilwell , $P_{IS}$, parameters. In the SME non-rotating resonator experiments in the laboratory test for four components of the $\tilde{\kappa}_{e-}$ tensor and three components of the $\tilde{\kappa}_{o+}$ tensor, with the scalar coefficient $\tilde{\kappa}_{tr}$ and $\tilde{\kappa}_{e-}^{ZZ}$ unmeasurable. Rotation in the SME enhances the sensitivity to the seven components and also allows the determination of the $\tilde{\kappa}_{e-}^{ZZ}$ component. We have shown that all resonator experiment exhibit the same relative frequency spectrum to the putative signal to within a multiplicative sensitivity factor, $\mathcal{S}$. This was utilized to compare the sensitivity of different FP and WG resonator configurations, leading to the proposal of some new dual-mode resonator experiments.

We applied the above analysis to our experiment at the University of Western Australia, which is based on rotating cryogenic sapphire whispering gallery mode microwave oscillators. In summary, we presented the first results of the experiment, which we set bounds on 7 components of the SME photon sector (Table \ref{tabb}) and $P_{MM}$ (Table \ref{Tabrms}) of the RMS framework up to a factor of 7.5
more stringent than those obtained from previous experiments. We also set an upper limit ($2.1(5.7)\times10^{-14}$) on the previously unmeasured SME component $\tilde{\kappa}_{e-}^{ZZ}$. To further improve these results, tilt and environmental controls will be implemented to reduce systematic effects. To remove the assumption that the $\tilde{\kappa}_{o+}$ and $\tilde{\kappa}_{e-}$ do not cancel each other, data integration will continue for more than a year.

\section{Acknowledgment}

This work was funded by the Australian Research Council.

%
%
\input{referenc}


\printindex
\end{document}

%% file: referenc.tex
%
%

%
%